\documentstyle[12pt,epsfig]{article}
\newcommand{\blankline}{\vskip .3cm}
\newcommand{\f}{\begin{equation}}
\newcommand{\ff}{\end{equation}}
\begin{document}
  \centerline{\LARGE  How far are we from the quantum theory of gravity?}
  \blankline
  \blankline
  \rm
  \centerline{Lee Smolin}
  \blankline
  \centerline{\it Perimeter Institute for Theoretical Physics}
  \centerline{\it Waterloo,  Canada}
  \centerline{\it  \ \ and }
  \centerline{\it Department of Physics, University of Waterloo}
  \centerline{lsmolin@perimeterinstitute.ca}

  \blankline
  \blankline
  \blankline
  \blankline
  \centerline{March 19, 2003}
  \blankline
  \blankline
  \blankline
  \blankline

  \centerline{ABSTRACT}

  An assessment is offered of the progress that the major
  approaches to quantum gravity have made towards the goal of
  constructing a complete and satisfactory theory. The emphasis is
  on loop quantum gravity and string theory, although
  other approaches are discussed, including dynamical triangulation 
  models (euclidean and lorentzian) regge calculus models, causal 
  sets, twistor theory, non-commutative geometry 
  and models based on analogies to condensed 
  matter systems. We proceed by listing the questions
  the theories are expected to be able to answer. We then compile two 
  lists: the first details the actual results so far achieved in each 
  theory, while the second lists conjectures which remain open.  
  By comparing 
  them we can evaluate how far each theory has progressed, 
  and what must still be done before each theory can be considered a 
  satisfactory quantum theory of gravity.   We find
  there has been impressive recent progress on several fronts. 
  At the same time, important issues about loop quantum gravity are so far 
  unresolved,
  as are key conjectures of string theory.  
  However, there is a reasonable expectation that experimental tests
  of lorentz invariance at Planck scales may in the near future make 
  it possible to rule out one or more candidate quantum theories of
  gravity.

  \blankline
  \blankline
  \eject
  \tableofcontents
  \eject
  \vfill

  \section{Introduction}

  \begin{quotation}

      {\it This paper is dedicated to Stanley Deser, 
      Bryce DeWitt, Cecille 
      Morette-DeWitt, David Finkelstein, Chris Isham, Karel Kuchar, 
      Roger Penrose
      and John Archibald Wheeler, each pioneers who  
      have been and who remain continuing
      sources of inspiration and encouragement for all of us working
      in quantum gravity.}

  \end{quotation}

 For most of the twentieth century physics proceeded with two
 fundamental physical theories, quantum theory and general
 relativity.  The latter is Einstein's theory of space, time and
 gravitation, while the former describes essentially everything else
 in nature. This situation was possible, because there were no experiments
 that probed regimes in which both quantum and gravitational effects
 were present.  At the same time, the fact that nature is one entity
 meant that there must eventually be discovered a unification of
 quantum theory and general relativity, which could stand as a single
 theory of nature.  Such a theory is called a quantum theory of
 gravity.

 Not so many years ago, it was common to hear the statement that
 there is no quantum theory of gravity and that the invention of such
 a theory is far off.  Although a few people have worked on the
 problem of quantum gravity since the 1950's, no great progress was
 made until the early seventies, apart from technical developments
 which ruled out various approaches\footnote{For a brief history 
 of research in quantum gravity, see \cite{carlo-history}.}.  
 These included standard perturbative approaches, which attempted
 to base quantum gravity on a Feynman perturbation theory for graviton modes,
 of the form,
 \f
 g_{ab}=\eta_{ab}+ h_{ab}.
 \label{naive}
 \ff
 Here $h_{ab}$ is defined to be a small 
 excitation on a flat background $\eta^{ab}$.
 All such approaches to the quantization of general relativity 
 were found to fail at some low order in perturbation
 theory, yielding theories that were perturbatively 
 nonrenormalizable. Various attempts were made to save the
 situation at the level of an expansion of the form of (\ref{naive})
 and they all failed. For example, one can add to the Einstein action 
 terms in the square of 
 the curvature; perturbative renormalizability is then accomplished, 
 but  at the 
 expense of perturbative unitarity. The same holds for attempts to 
 resolve the problem by adding new 
 degrees of freedom, such as dynamical torsion or 
 non-metricity. In each case one can construct theories that are 
 perturbatively unitary and theories that are perturbatively 
 renormalizable, but not theories that have both properties.  Various
 attempts were made to construct alternative expansions 
 such as $1/N$ expansions, $1/D$ 
 expansions, the Lee-Wick mechanisms 
 etc.\footnote{If the reader doesn't know what these are don't 
 worry, they didn't work!}; they all 
 suffered the same fate. 
 There was a brief period of excitement about supergravity, but after
 a while it was realized that all supergravity theories are likely to
 suffer the same fate when treated perturbatively. 

 There were, nevertheless, significant advances in the 1970's. 
 Around 1971
 several striking results were found, concerning the behavior
 of quantum fields on a few spacetime backgrounds besides Minkowski
 spacetime. These included
 Bekenstein's discovery of the entropy of black holes\cite{bek1}, Hawking's
 discovery that black holes are hot\cite{hawking}, and radiate,
 and Unruh's discovery
 that even the vacuum of flat spacetime behaves as a thermal state
 when viewed by an accelerating observer\cite{unruh}.  These effects all point
 to the possibility of a deep connection between spacetime, quantum
 theory and thermodynamics, which has fascinated people ever since.

 Still, this was not quantum gravity, as the geometry of spacetime,
and the gravitational field, were still treated as in Einstein's
classical theory. Real, undeniable, progress on quantum gravity
began only in the mid 1980's. The reason was the almost simultaneous
invention of two approaches to quantum gravity which each quickly achieved
impressive advances towards the solution of some aspect of quantum gravity.
These two developments were string theory\cite{GSW,Joe} and
loop quantum gravity\cite{carlo-review}-\cite{volume}.

Since then both string theory and loop quantum gravity have been the
subject of a large and intense effort by many people\footnote{For
popular accounts of string theory and loop quantum gravity,
see \cite{elegant} and \cite{threeroads}.  Further pedagogical material
is available on several websites\cite{web}.}. After 18
years, a large number of results have accumulated about each theory.
In addition, in recent years several new approaches have been invented,
the include causal dynamical triangulations, non-commutative geometry,
causal sets and approaches based on analogies to condensed matter physics.
The main purpose of this essay is to make an evaluation of where each
theory stands in relation to the main questions that a quantum theory
of gravity is expected to answer.

One reason to carry out such an evaluation is that, while the
undeniably impressive progress on several sides has generated a lot of
excitement among both experts and the wider community, there appears
at the same time
to be a great deal of confusion about exactly what each of the
theories has so far achieved.  This is perhaps surprising, as it does
not appear to have been the case with earlier theoretical triumphs
such as quantum theory or relativity.  Still, one only has to talk to
a wide enough selection of experts to get the impression that there is
quite a lot of disagreement about the significance of the results so
far achieved on each side\footnote{See, for example, \cite{controversy}.
For a critical view of string theory by one of its pioneers, see,
\cite{freidan}.}.  In some cases there is even confusion about what
the actual results are.

This confusion has several sources.  The first is
the gap that necessarily exists between the highly technical and qualified
language that must be used to describe the actual results and the
more general language that is used to convey their  significance
to a wider audience, not only to non-scientists, but to physicists and
mathematicians who are not experts in the theory in question.  It is
also unfortunately true that some, although of course not all,
proponents of each theory have sometimes simplified the statements of results
in presentations for non-experts in such a way as to appear to claim
results which have in fact not yet been shown.
There is also a lot of confusion caused
by the fact that in a few crucial cases, there are conjectures
which are widely believed by experts, in spite of the fact that
they remain unproven. Additional confusion comes from the fact that
some of these conjectures come in different inequivalent forms.

Another source of confusion is the very unfortunate isolation in
which each community works.  It is striking that there has never been
a talk on loop quantum gravity at one of the annual string theory
conferences.  And, while there are ongoing conversations between some
people in the two communities, there are very few people
who have done technical work in both theories.  As a result, many experts
in one approach have only a very superficial understanding of the
other.

The sad result is
that many members of each community sincerely believe that
the approach they work on is the only viable approach to quantum
gravity.  This of course causes still more alienation which further
separates the two communities. It is very offensive for someone
working on loop quantum gravity to listen to a talk or read a book or
paper that begins, as they unfortunately often do,
with the assertion that ``string theory is the only quantum theory of gravity.''
At the same time, string theorists listening to talks on loop quantum
gravity are often puzzled by the lack of interest in supersymmetry
and higher dimensions, which string theory has shown seem to be
required to satisfy certain criteria for a good theory\footnote{In
fact there are results that show that loop quantum gravity extends
easily to supergravity, at least through $N=2$ and there are even
partial results on $11$ dimensional
supergravity\cite{yime-holo,superyi,super,11d}.
Moreover
some results on spin foams extend to $d>4$\cite{higher}.}.

For all of these reasons, it seems important to attempt to carry out
an objective evaluation of the status
of these two research programs.  The present paper is an attempt to do
just this.  It began, indeed, as a personal project, for as someone
who has worked on both theories I found myself in a situation in
which I was quite confused and  puzzled about the status of each of the
two theories. In order to decide which to continue working in, and on
what problems, I decided to undertake an analysis of the actual
results in each case.  By doing so one can see more clearly what
would have to be done in each case to move the theory from its present
situation to the status of a true physical theory. 

\subsection{Methodology}

The method I decided to pursue is mirrored in the structure of the
paper that follows. I began by writing down a list of the
questions that the theories are expected to be able to answer.
Then I wrote down as complete a list as I could of the major results
of the two theories.  I also made a list of the conjectures that
have been made in the course of the development of the two theories,
and for each considered whether it had been demonstrated, or
disproven, by the results.  Or, if neither was the
case, I tried to establish to what extent the actual results could be taken
as providing evidence for the conjectures.

There are different standards by which physicists and mathematicians
judge the reliability of results. I took as appropriate those of
mainstream theoretical
physics.  I do not require the rigor of mathematical physics,
although, as will be pointed out, there are results for each theory
that are at this level.
When a result is claimed regarding a quantum field theory, it
should have been obtained in a context in which all expressions have
been regulated, all divergences or ambiguities are resolved, and careful
attention has been given to technical issues such as how the gauge invariances
of the classical theory are maintained in the quantum theory.  When a
path integral is involved it should be fully defined in
terms of a well defined measure, or else expressed as a discrete summation.

This led in each case to two lists, the first of results, the second
of conjectures and open issues.
These are summarized in Table 1, which indicates the extent to which
each of the theories answers each of the questions posed.
After this I asked what steps remain
before each theory
might be considered complete. By this I meant that it is precisely
formulated and well understood mathematically and conceptually, that
there are methods to carry out calculations leading to predictions
for real experiments, and that at least a few experiments have been
done which either support or falsify the predictions of the theory.

Table 1 and the lists of things still to do comprise the main
conclusions of this essay.  What to do about the present situation,
whether to continue to investigate one or the other or both theories
is a matter of opinion and an individual's research strategy.  I will
indicate my own personal conclusions at the end, however I want to
stress that I do not believe that all experts will, or even should
agree on these questions. Science works best when there is a variety
of viewpoints investigated, and when there is room in the community
for people who investigate a range of viable approaches to any
unsolved problem.  But I do think
that it is a useful exercise to try at least to come to a consensus
about what the evidence is, what has been done and what remains to be
done. I hope that this essay will contribute to that goal.

Let me describe some issues which arose in carrying out this program.
First, it is important to distinguish two kinds of
results. The first are  results which further our
  understanding of physical questions the theory was originally invented
to answer. These are to be distinguished from
 results which resolve issues and puzzles raised by the theory
  itself, whose solution will help us understand the theory better,
  but which will not lead to the answer to a question about nature.
  Results of the first kind we may call {\it substantial},
whereas results of the second kind can be called {\it internal.}

  While this is not an ironclad distinction it is a useful one.
  A lot of progress can be (and often must be) made on understanding
  the mathematical structure of a theory, without any actual progress
  being made on any question about the natural world. In evaluating the
  status of a theory we may be impressed by progress on the second
  kind of question, but the main focus must be  the first kind.

  This is especially the case with a complicated theory like string
  theory, which has many, perhaps an infinite number of,  versions of
  which do not describe the universe we live in.  In a case like this
  we must distinguish between measures of activity, which may result
  in various aspects of the theory being explored which do not
  relate, even indirectly,
  to nature, and measures of progress  leading to
  understanding some feature of the natural world or to  new predictions
  for real, doable experiments.

 To distinguish between these two kinds of questions
  it is important to keep in mind what aspects of
  nature are known from experiment and which are postulated by theory.
   If a  result addresses a problem raised by believing in some fields,
symmetries or dimensions for which there is so far no observable evidence,
then it is an internal result.

  Thus, among the many published results, I have included here mostly
  substantial results. I have included internal results when they
  are important to judge the likelihood of the truth of the central
  conjectures of the theories.

  Within the class of substantial questions we may make some
  distinctions according to subject.  The focus of interest in this
  essay is on
the questions that any quantum theory of gravity must answer.
A second set of questions comes from cosmology. They arise from the
existence of puzzles arising concerning cosmological data that appear to
have no solution except in unknown effects at the Planck scale.
While it is not necessary that a quantum theory of gravity answer
these cosmological questions, there is still, because of this,
a good possibility that a
quantum gravity theory may make testable predictions about
cosmological data. This is reason enough to include these questions in
the present evaluation.

There are of course also questions about elementary
particle physics. Here the two theories are in rather different
situations. String theory claims to be a unified theory of all
interactions, hence it must be judged on its ability to make verifiable
predictions about the elementary particles. Loop quantum gravity makes
no claim to be other than the quantum theory of gravity, and in fact
appears able to incorporate equally well a wide variety of matter
fields and interactions.  So while loop quantum gravity can easily
incorporate the standard model of particle physics, it, at least so
far, makes no claims to explain any features of the standard model.

Here string theory has a big potential
advantage. Given the fact that it is truly a unified theory, were it
to make striking and unique predictions for elementary particle
physics that were confirmed experimentally, this would be strong reason
to believe in
string theory. At the same time, this is also a potential vulnerability,
for if it makes no such predictions it looses credibility.

There is here a real difference between the two approaches.  There is
no {\it a priori} reason that the problem of quantum gravity is
strongly linked to the question of unification.  After all the quantum
theory of electromagnetism, $QED$, has little to say about unification and
does not strongly constrain the matter degrees of freedom or what
other interactions there are in nature. At the most we can say that to
eliminate a potential inconsistency at high energy-called the Landau
ghost- $QED$ should be imbedded in an asymptotically free gauge
theory. But there are many of those, and even this does not imply the
unification of all the gauge forces.

Nor is there an absolutely {\it compelling} reason
to believe in a unification of gravity with the other forces. Gravity
plays a unique role in physics, as it is connected with the geometry of
space and time.  Thus, it is only
gravity that can be understood to be a consequence of the fact that the
specification of inertial frames is local and dynamically determined.
It is of course possible that, as has been proposed for decades, the
other interactions also come from the dynamics of spacetime geometry,
such as the curvature of extra dimensions. However, while this is an
extremely
attractive idea, it must also be admitted that there is so far no compelling
argument from  experiment or theory for either the existence of the
extra dimensions or the necessity that the other forces be described
in terms of them.

The best evidence that the problem of quantum gravity is related to the
problem of unification comes instead from perturbation theory. It
comes first from the fact that supersymmetry appears to be required to
have a perturbative quantum theory that
includes gravitons and is also exactly lorentz invariant.  Further,
among the possible supersymmetric gravitational theories, we can
make a good
case for the likelihood of complete consistency only in the case of
the string theories.
This is consequential, and is a strong argument for taking string
theory seriously, at the very least as an effective description of
a fundamental theory, good at scales less than the Planck scale.
But it could still be wrong, for example, it could be that Lorentz
invariance is broken or modified at Planck
scales\cite{GAC1}-\cite{testreviews}.
Were this to be
discovered experimentally (and, as we will mention below, there are
experimental results that may be interpreted as indicating a failure
of Lorentz invariance)\cite{AC-Piron}
not only would string theory be not needed,
but one of its main assumptions would have been falsified.

Finally, there are questions about foundational issues concerning
quantum theory and the nature of time.  The situation here is similar
to that of unification. Good arguments have been put forward by
several of the deepest thinkers in the field-people like
Roger Penrose\cite{Roger-books} and
Gerard `t Hooft\cite{thooft}-that
the problem of quantum gravity cannot be solved
without revising the principles of quantum theory.  But there is no
experimental evidence for such modifications and it remains possible
that such arguments are wrong and that quantum gravity, like quantum
electrodynamics, can be solved without forcing a deepening of our
understanding of the principles of quantum theory.

One reason to side with the deep thinkers is the difficulty of
formulating quantum theory sensibly in a cosmological context in which
the observers must be part of the
system\cite{Fotini-Wheeler,LOTC,threeroads}.  But still, it may be that
the problem of discovering the quantum theory of the gravitational
field in local regions of spacetime may be solved separately, while
the problems of quantum cosmology remain open for smarter people in the future
to finally resolve.

After listing the questions the theories may aspire to answer, I  give
a quick survey of the similarities and differences between the two
theories. Indeed, it is striking and, I believe, non-trivial, that
the two theories have a lot in common, so much so that any evaluation
of their future must take into account the possibility that they will
turn out to be different sides of a single theory.  At the same time,
there are big differences between them, and some of these can be
recognized immediately.  After this, we begin the detailed listing of
results and open conjectures for each theory.

  Before closing the introduction I should state my own situation
  with respect to the two theories.  Since 1984 I have worked on both string
  theory and loop quantum gravity. If I have so far contributed more to loop
  quantum gravity, the majority of my papers since 1998 concern string or
  $\cal M$ theory. I have also given graduate courses in both string
  theory and loop quantum gravity, I've had Ph.D. students and
  hired postdocs working in both areas and I attend conferences
  in both areas. So I think I do know both of them in enough
  technical detail to attempt this kind of evaluation.
  In particular,  I have tried to make my own choices
of which program to work on based on an objective evaluation of their
potential to solve the key questions in quantum gravity. And, as
the theories developed,
I have made this choice differently at different times in the
last 18 years.

Of course, I do not expect everyone will be happy with the conclusions I
reach here. I myself was surprised by the conclusions I was led to by
going through the exercise of writing this paper, and they have
changed my own research priorities.
But I do believe that any honest person who takes the time
to acquaint themselves with the actual technical
details of each theory sufficiently to understand the detailed
statements of assumptions and results, will, if they reflect carefully
on the actual evidence at hand, and if they are in fact open enough to accept
any conclusion the evidence supports, reach essentially
the same conclusions I do here.

As we proceed, I will lay out my conclusions with care, and with suitable
attention to careful statements of assumptions and results. I am more
than happy to discuss any of the conclusions I reach with anyone, and
I am open to having my views changed, either by someone explaining
something I missed or misunderstood, or, of course, by new results.

\subsection*{Other approaches}

Before going on it is important to mention that string theory and loop
quantum gravity are not the only approaches to quantum gravity
that have been invented and studied. Other approaches include causal
sets\cite{causalsets},
dynamical triangulations\cite{dynamical},
causal dynamical triangulations\cite{AL},
twister theory\cite{twisters},
non-commutative geometry\cite{NCG}, supergravity, approaches based
on analogies to condensed matter physics, etc. Each of these is
motivated by rather compelling arguments, and each has been pursued
vigorously by a community of smart people.   Several of them, such as
dynamical triangulations and causal dynamical triangulations, have
achieved very significant results.

While none of these approaches has gained nearly the number of results found
for string theory or loop quantum gravity, some of them do nevertheless address
key issues and so deserve mention in any survey of progress in quantum gravity.

It is also the case
that some aspects of some of these approaches have been incorporated
into string theory or loop quantum gravity. For example,
non-commutative geometry appears in both, and causal sets play a role
in loop quantum gravity.  To further complicate the situation,
some approaches can, if one wishes, be considered to be
subcases or limits of string theory or loop quantum gravity, but may
also stand
on their own.  For example supergravity can be considered to be a
limit of string theory, although a few purists may want to insist
that there still may be a quantization of supergravity which is not a
string theory.  Similarly, dynamical triangulation models can
be considered to comprise a class of loop quantum gravity models, and
the methods used to study them likely extend to general loop quantum
gravity models. But there is no necessity to consider them as loop
quantum gravity models.

\section{Physical questions the theories should answer}

  \subsection{Questions concerning quantum gravity}

  We begin with the problems of quantum gravity itself.
  The correct quantum theory of gravity must:

  \begin{enumerate}

  \item{}{\it Tell us whether the principles of general relativity
and quantum mechanics are true as they stand, or are in need of modification.}

  \item{}{\it Give a precise description of nature at all scales,
  including the Planck scale.}

  \item{}{\it Tell us what time and space are, in language fully
compatible with both quantum theory and the fact that the geometry of
spacetime is dynamical. Tell us how
light cones, causal structure, the metric, etc are to be described
quantum mechanically, and at the Planck scale.}

  \item{}{\it Give a derivation of the black hole entropy and
temperature. Explain how the black hole entropy can be understood as a
statistical entropy, gotten by coarse graining the quantum description.}

  \item{}{\it Be compatible with the apparently observed positive,
but small, value of the cosmological constant. Explain the entropy of
the cosmological horizon.}

  \item{}{\it  Explain what happens at singularities of  classical
general relativity.}

  \item{}{\it Be fully background independent. This means that no
classical fields, or solutions to the classical field equations appear in the
theory in any capacity, except as approximations to quantum states and
histories. }

  \item{}{\it Predict new physical phenomena, at least some of which
are testable in current or near future experiments.}

  \item{}{\it Explain how classical general relativity emerges in an
  appropriate low energy limit from the physics of the Planck scale.}

  \item{}{\it Predict whether the observed global lorentz invariance
of flat spacetime is realized exactly in nature, up to infinite boost
 parameter, or whether there are modifications of the realization of
lorentz invariance for Planck scale energy and momenta.}

  \item{}{\it Provide precise predictions for the scattering of
gravitons, with each other and with other quanta,
 to all orders in a perturbative expansion around the semiclassical
 approximation.}

\end{enumerate}

  These are a lot of questions, but it is hard to imagine believing in
  a quantum theory of space and time that did not answer each one.
 However, there is one that cannot be over-emphasized,
 which is the requirement of background independence.   There are two
 reasons for making this requirement. The first is a matter of
 principle.  Over the whole history of physics, from the Greeks
 onwards, there have been two competing views about the nature of
 space and time. The first is that they are not part of the dynamical
 system, but are instead eternally fixed, non-dynamical aspects of the
 background, against which the laws of nature are defined.  This was
 the point of view of Newton and it is generally called the {\it
 absolute} point of view. The second view holds that the geometry of space
 and time are aspects of the dynamical system that makes up the
 universe. They are then not fixed, but evolve as does everything else,
 according to law. Further, according to this view, space and time are
 {\it relational}.  This means there is no absolute meaning to where or
 when an event occurs, except as so far as can be determined by
 observable correlations or relations with other events. This was the
 point of view of Leibniz, Mach and Einstein and is called the {\it
 relational} point of view.

 Einstein's theory of general relativity is an instantiation of the
 relational point of view. The observations that show that gravitational
 radiation carries energy away from binary pulsars in two degrees of
 freedom of radiation, exactly as predicted by Einstein's theory, may
 be considered the experimental death blow to the {\it absolute} point
 of view.  The fact that two, and not five, degrees of freedom are
 observed means that the gauge invariance of the laws of nature
 includes spacetime diffeomophism invariance. This means that the
 metric is a completely dynamical entity, and no component of the
 metric is fixed and non-dynamical.

 As argued by Einstein and many others since, the
  diffeomorphism invariance is tied directly to the background
  independence of the theory. This is shown by the hole
  argument\cite{hole}, and
  by Dirac's analysis of the meaning of gauge symmetry\cite{Dirac-book}.
  There are good discussions of this by
  Stachel\cite{stachel}, Barbour\cite{julian-gr}, Rovelli\cite{carlo-gr}
  and others\cite{threeroads,LOTC}.

 Thus, classical general relativity is background independent.  The
 arena for its dynamics is no spacetime, instead the arena is the
 configuration space of all the degrees of freedom of the gravitational
 field, which is the metric modulo diffeomorphisms.

 Now we can ask, must the quantum theory of gravity also be
 background independent? To have it otherwise would be as if some
 particular classical Yang-Mills field was required to {\it define}
 the quantum dynamics of $QCD$, while no such fixed, non-dynamical
 field need be specified to define the classical theory.  Still a
 number of people have expressed the view that perhaps the quantum
 theory of gravity requires a fixed non-dynamical spacetime
 background for its very definition.  This seems almost absurd, for it
 would mean taking some particular solution (out of infinitely many)
 to the classical theory, and making it play a preferred role in the
 quantum theory. Moreover, there must be no experimental way to
 discover which classical background was taken to play this preferred
 role, for if any effect which depended on the fixed background
 survived in the low energy limit, it would break diffeomorphism
 invariance. But this would in turn mean that diffeomorphism
 invariance was not an exact gauge symmetry in the low energy limit,
 and this would imply that more than two degrees of freedom of the
 metric would be excited when matter accelerated. But this would
contradict the extreme sensitivity of the agreement between general
 relativity and the rate of decay of binary pulsar orbits.

 Thus, arguments from both principle and from experiment reinforce the
 conclusion that nature is constructed in such a way that, even in the
 quantum domain, all the degrees of freedom of the spacetime geometry
 are dynamical.  But if this is the case no fixed classical metric can
 play any role in the formulation of the quantum theory of 
 gravity\footnote{It is
 sometimes argued in rebuttal that an acceptable theory may be 
 formulated in such a way
 that the quantum theory depends on a classical background, but any 
 of a large number of
 backgrounds may be used, so that the theory does not require one 
 special background.  This
 misses the point, as such a theory in fact consists of a long list 
 of quantum theories, one for
 each background. This fails to realize the idea that quantum spacetime 
 as a whole is dynamical, so that
 the different backgrounds arise as solutions of the quantum dynamics. 
 It is not enough
 that the different backgrounds may be solutions of 
 different {\it classical} equations, for
 that leads to a mixed and most likely inconsistent theory in 
 which the geometry is split in such
 a way that one part (the background) solves a classical equation, 
 while the other part (the
 gravitational waves ``on the background" satisfy quantum equations 
 that depend on the choice
 of background. Such an approach may arise as an approximation to a 
 fundamental theory, but it
 cannot be a fundamental theory in itself.}.

   \subsection{Questions concerning cosmology}

   Next we mention cosmological puzzles that are so far unsolved and
   that are widely believed to require Planck scale physics for their
   resolution.

  \begin{enumerate}

  \item{}{\it Explain why our universe apparently began with
extremely
  improbable initial conditions\cite{Roger-books}.}

  \item{}{\it In particular, explain why the universe had at grand
unified
  times initial conditions suitable for inflation to occur or,
  alternatively, give an alternate mechanism for inflation or a
mechanism by
  which the successes of inflationary cosmology are duplicated.}

  \item{}{\it Explain whether the big bang was the first moment of
time, or whether there was something before that.}

  \item{}{\it Explain what the dark matter is.  Explain what the dark
energy
  is. Explain why at present the dark matter is six times as dense as
  ordinary hadronic matter, while the dark energy is in turn twice as
dense
  as the dark matter.}

  \item{}{\it Provide predictions that go beyond those of the
currently
  standard model of cosmology, such as corrections to the CMB spectra
  predicted by inflationary models.}

\end{enumerate}

  \subsection{Questions concerning unification of the forces}

  Next, we mention problems in elementary particle physics that
  must be resolved by any unified theory of all the interactions. As
  string theory must, if true, be such a theory, it must be evaluated
  against progress in answering these questions. It is also possible,
  but not as necessary, that loop quantum gravity offer answers to
  some of these questions.

  \begin{enumerate}

  \item{}{\it Discover whether there is a further unification among
the forces, including gravity or not.}

  \item{}{\it Explain the general features of the standard model of
  elementary particle physics. i.e. explain why the forces are
described by
  a spontaneously broken gauge 
  theory with group $SU(3)\times SU(2)\times U(1)$, 
  with fermions in the particular chiral representations
observed.}

  \item{}{\it Explain why there is observed a large hierarchy in the
ratio
  of masses observed, from the Planck mass, down to the neutrino
masses and finally down to the cosmological constant.  Discover the
mechanism by which the hierarchy was created, whether
by spontaneous breaking of a more unified theory or by other means.
Explain why the cosmological constant is so small in Planck units.}

  \item{}{\it Explain the actual values of the parameters of the
standard model: masses, coupling constants, mixing angles etc. Explain
the observed value of the cosmological constant.}

\item{}{\it Tell us whether there is a unique consistent theory of
nature that implies unique predictions for all experiments or whether, as
has been sometimes proposed, some or
all of the questions left open by the standard model of particle physics
are to be answered in terms of choices among possible consistent
phenomenologies allowed by the fundamental theory.}

\item{}{\it Makes some experimental predictions for phenomena that are
unique to that theory and which are testable in present or near future
experiments?}

  \end{enumerate}

  \subsection{Foundational questions}

  Finally, there are the questions in the foundations of quantum
  theory, which many people believe are closely related to the
  problem of quantum gravity.

  \begin{enumerate}

 \item{}{\it Resolve the problem of time in quantum cosmology}

  \item{}{\it Explain how quantum mechanics is to be modified to
apply to a closed system such as the universe that contains 
its own observers.}

\item{}{\it Resolve the puzzle about where the 
information apparently lost in black hole evaporation goes.}

\end{enumerate}

  \section{A tale of two theories}

  Before listing the main results and open issues of each theory, it
  is useful to survey the main points in common, and main differences
  of the two theories.  Both the similarities and differences are
  striking and non-trivial, and it is probably useful for the reader
  if they are highlighted here, before we are involved, necessarily,
  in a great deal of details and fine distinctions necessary to reach
  a careful evaluation of each theory.

  \subsection{Common postulates}

  String theory and loop quantum gravity are each a development of a
set of ideas originally introduced in the 1960's to understand hadronic
physics.  As such they share some common postulates.

  \begin{itemize}

  \item{}{\bf The fundamental theory is not a conventional Poincare 
  invariant local field theory.}

  \item{}{\bf The fundamental excitations are extended objects.}
These include one dimensional excitations and two
(and perhaps higher dimensional) membrane-like excitations.

  \item{}{\bf Duality}  The one dimensional excitations have a dual
  description as quanta of electric flux of a non-abelian gauge
theory. The  higher dimensional excitations have a dual description in terms of
higher dimensional electric and magnetic fluxes.

  \item{}{\bf The holographic principle.}  This is a recently proposed
  principle which, if true, is the first principle we have that uniquely
  concerns the quantum theory of gravity. It says, roughly, that
  observables for quantum gravity theories in $d$ spacetime dimensions
  can be evaluated in terms of data on $d-1$ dimensional
  surfaces\cite{louis-holo}-\cite{weakstrong}\footnote{The
  principle 
  was independently proposed by Louis Crane\cite{louis-holo} and
  Gerard 't Hooft\cite{thooft-holo} in 1993. It was applied to 
  string theory by Susskind in \cite{lenny-holo} where it inspired
  developments such as the AdS/CFT conjecture.  Crane's formulation
  inspired key developments in loop quantum gravity including the discovery
  that Chern-Simons theory plays a role in describing the quantum
  geometry of boundaries and horizons, leading to the present
  understanding of black hole entropy\cite{linking,isolated}.}.
  These surfaces may be boundaries of the spacetime or, in the
  cosmological case, may be surfaces embedded in the spacetime.

  Different versions of the holographic principle have been proposed,
  which differ on the extent to which the theory can be completely
  reduced to a dynamical theory on the lower dimensional surface.
  For more details see\cite{weakstrong}.

  \end{itemize}

  The fact that string theory and loop quantum gravity share these
  common postulates is reflected in the fact that the mathematics
  employed in their formulation overlaps. For example, they both
  employ conformal field theory and the representation theory
  of quantum groups.   Both theories may be
  formulated in a language in which all the degrees of freedom are
  represented as large matrices. These formulations are
  non-perturbative in the sense that the dynamics of the matrices
  code an infinite number of terms in a perturbation
  theory\footnote{For the matrix formulation of loop quantum
  gravity see\cite{MLQG}, for string and $\cal M$ theory formulated
  as a matrix model see \cite{BFSS,IKKT}}.

  However there are very significant differences as well.

\subsection{Why string theory and loop quantum gravity differ}

String theory and loop quantum gravity both begin by taking the one
dimensional extended objects,  which by duality correspond
to electric flux lines of
a quantized gauge field, to be fundamental degrees of freedom of the
theory.  They differ in two ways:

\begin{itemize}

    \item{}{\bf Difference one.}  {\it The strings are taken to move in
a classical background characterized by a fixed choice of a metric
and other   classical fields. The loops are taken to exist at a more
    fundamental level, at which there are no classical metrics or
    other fields.}

    \item{}{\bf Difference two.}  {\it The gauge field in the case of
    loops is taken to gauge all or part of the local lorentz
    transformations.  The gauge field in the case of open strings is
    taken to correspond to a Yang-Mills field.}

    \item{}{\bf Difference three.}
    {\it  The two approaches take very different strategies to
    address the failure of general relativity to exist as a
    perturbatively renormalizable quantum field theory.  These have
    to do with the attitude to the physical assumptions that underlie
    the use of perturbation theory. These postulates include i)
    spacetime is smooth down to arbitrarily small scales, so that
    there are linearized perturbations of arbitrarily short
    wavelength. ii) Global lorentz symmetry is an exact symmetry of
    the spectrum of fluctuations around the quantum state
    corresponding to Minkowski spacetime, good to arbitrarily small
    wavelength and large boost parameter.

    These two postulates are assumed by string theory to be exact. The
    attitude taken is to search for a perturbative theory
    incorporating gravitons in which they can be exactly realized.

    In contrast, loop quantum gravity takes the attitude that we must
    make a quantization of general relativity that does not rely on
    these two assumptions. Indeed, as global lorentz invariance is not a
    symmetry of classical general relativity, it cannot be assumed
    in any exact quantization of the theory. These two 
    assumptions are then to be tested, in the sense that one must see to
    what extent they are recovered in the classical limit of the
    quantum theory.
    In fact, as we shall see, the evidence is that they are false
    in at least one consistent quantization of general relativity.}

    \end{itemize}

  As a result of these differences the two theories have different
  postulates.  They lead, as well, to very different
  physical pictures. Consequently, the  two theories make quite
  different predictions for future experiments.  It is
  worth mentioning these at the beginning.

  \subsection{Characteristic predictions of string theory}

  String theory appears to require that the world have large numbers of
  so far unobserved
  dimensions, degrees of freedom and symmetries\cite{GSW,Joe,elegant}.
  While we will discuss
  this in detail below, it can be said that string
  theory requires that nature have 6 or 7 dimensions of space beyond
  the three that are observed. It  also predicts the existence of a
  new kind of symmetry, called supersymmetry, which is also so far
  unobserved. This is a symmetry that relates fermions to bosons.
  Unfortunately, it appears that supersymmetry cannot be used to relate
  any of the presently known fermions to any of the presently known
  bosons. Thus, supersymmetry, and string theory, predict that there
  are a great many unobserved elementary particles.

  Two things must then be said. There is so far no evidence at all
  from observation
  for any of the additional dimensions, symmetries or particles that
  string theory predicts\footnote{There are some facts that are taken 
  as possible indirect 
  evidence for supersymmetry in particle physics. One has to do 
  with the question of whether the gauge and Yukawa coupling constants
  meet at a single grand unification scale. There is approximate but not 
  exact unification in the standard model. The unification is closer 
  in the minimal supersymmetric standard model,
  in that the triangle made by the three running coupling
  constants is smaller and it is more plausible that unification is
  achieved by threshold effects\cite{raman}. However,
  the running of the coupling constants may also be influenced 
  by other factors such as neutrino masses\cite{neutrino}.}.  
  Second, string theory is not unique in
  predicting any of these features. String theory was preceded by the
  study of higher dimensional theories and ordinary theories with
  supersymmetry. These theories continue to be studied independently of
  string theory.  It is not easy to point to a doable
  experiment that would confirm a
  prediction of string theory, uniquely, that is not also a
  prediction of an ordinary supersymmetric or higher dimensional
  field theory.

  There is one assumption that string theory makes which is subject
  to experimental test.  This is that special relativity holds, at all
  distance scales, in its original form given by Einstein. In
  technical language this means that the theory assumes that lorentz
  invariance is an exact symmetry of the world we live in, neglecting
  only effects due to the curvature of spacetime.

  \subsection{Characteristic predictions of loop quantum gravity.}

  Loop quantum gravity also leads to characteristic predictions of new
  phenomena, but of a rather different type. In fact loop quantum
  gravity is completely compatible with the postulate that the world
  has only three spatial dimensions, and one time dimension, and
  is known to be compatible with a large range of assumptions about
  the matter content of the world, including the standard model.
  So it does not require
  any dimensions, symmetries or degrees of freedom beyond what are
  observed. At the same time, there are versions of loop quantum
  gravity that incorporate supersymmetry (at least up to $N=2$,)
  and many results extend to higher dimensions. So were experimental
  evidence for either supersymmetry or higher dimensions this
  would not pose a problem for loop quantum gravity.

  Instead, the predictions of loop
  quantum gravity concern the structure of space and spacetime at very
  short distances. In particular, loop quantum gravity predicts that
  the smooth picture of spacetime in classical general relativity is
  actually only a coarse grained approximation to a discrete
  structure, in which surfaces and regions can have only certain,
  discrete quantized values of areas and
  volume\cite{spain,volume,renate-volume,volume2,sn1}.
  Loop quantum gravity
  makes specific predictions for the discrete quantum geometry at
  short distances. Furthermore, these predictions are derived from
  first principles, hence they are not adjustable. In this way loop
  quantum gravity is different from previous approaches which postulate
  some form of discrete structure as a starting point, rather than
  deriving it as a consequence of the union of quantum theory and
  general relativity.

  It turns out that this has consequences for the question of whether
  special relativity, and lorentz invariance, is exactly true in
  nature, or is only an approximation which holds on scales much
  longer than the Planck scale\cite{GAC1}-\cite{testreviews}.
  Several recent calculations, done with
  different methods\cite{GP}-\cite{positive},
  yield predictions for modifications to the energy
  momentum relations for elementary particles.
  These are of the form,
\f
E^{2} = p^{2}+ M^{2} + \alpha l_{Pl}E^{3}+ \beta l_{Pl}^{2}E^{4}
+ ...
\label{modified}
\ff
where predictions have been found for the leading coefficients $\alpha$,
which generally depend on spin and helicity\cite{GP}-\cite{positive}.

This is then an area of disagreement
  with string theory.  Further, these modifications appear to be
  testable with planned experiments\cite{GAC1,AC-Piron,seth,testreviews}.
  Hence the different
  predictions of string theory and loop quantum gravity concerning
  the fate of lorentz
  invariance offer a possibility of experimentally distinguishing
  the theories in the near future.

  \section{The near term experimental situation}

The most important development of the last few years in quantum gravity
is the realization that it is now possible to probe Planck scale physics
experimentally.   Depending on dynamical assumptions
there is now good experimental sensitivity to the $\alpha$
terms in (\ref{modified}) for photons,
electrons and protons.  Increased sensitivity is expected over the next few
years from a number of other experiments so that it is not impossible
that even if the leading order $E^{3}$ terms are absent, it will
be possible to put order unity bounds on $\beta$, the coefficient
of the $E^{4}$ term.

However it is crucial to mention that to measure $\alpha$ and $\beta$
one has to specify how lorentz invariance is treated in the theory. There
are two very different possibilities which must be distinguished.

\begin{itemize}
    
\item{}{\bf Scenario A)} The relativity of inertial frames is broken and there exists a 
preferred frame.
In this case the analysis has to be done in that preferred frame. 
The most likely
assumption is that the preferred frame coincides with the
rest frame of the cosmic microwave background.  In such theories energy and
momentum conservation are assumed to remain linear.

\item{}{\bf Scenario B)} The relativity of inertial frames is preserved, 
but the lorentz transformations
are realized non-linearly when acting on the energy and momentum eigenstates of
the theory. Such theories are called modified special relativity or 
doubly special
relativity. Examples are given by some forms of non-commutative
geometry, for example, $\kappa-$Minkowski spacetime\cite{kappa}.  
In all such theories energy and momentum conservation 
become non-linear which,
of course, effects the analysis of the experiments.
In some, but not all, cases of such theories, the geometry of 
spacetime becomes
non-commutative.

\end{itemize}

Among the experiments which either already give sufficient
sensitivity to measure $\alpha$ and $\beta$, or are expected to by 2010 are,
\begin{enumerate}

        \item{}There are apparent violations of the GZK bound
        observed in ultra high energy cosmic rays (UHECR) detected
        by the AGASA experiment\cite{AGASA}. The experimental situation is
        inconsistent, but the new AUGER cosmic ray detector, which
        is now operational, is expected to resolve the situation
        over the next year or two. If there is a violation of the
        GZK bound, a possible explanation is Planck scale physics
        coming from (\ref{modified}) \cite{AC-Piron}.  
	
	In Scenario A) violations of
	the GZK bounds can be explained by either $E^{3}$ or
	$E^{4}$ terms in the proton energy-momentum relation.
	However, in case B) it is less natural
	to explain a
	violation of the GZK bounds by means of a Planck scale
	modification of the energy-momentum relations, but there
	are proposals for forms of such theories that do achieve this. 

       \item{}A similar anomaly is possibly indicated in Tev
       photons coming from blazers\cite{blazers}.  Similar remarks
       apply as to the explanatory power of Scenarios A) and B) in the 
       event that the anomaly exists. 

        \item{}A consequence of (\ref{modified}) is an energy
        dependent speed of light. This effect can be looked for in
        timing data of gamma ray busts. Present data bounds
        $\alpha < \approx 10^4$ \cite{wavelets} 
	and data expected  from the GLAST
        experiment is expected to be sensitive to $\alpha$ of
        order one in 2006 \cite{glast}. Note that this applies to both
	Scenarios A) and B). 

        \item{}Present observations of synchrotron radiation in the 
	Crab nebula, together with reasonable astrophysical assumptions,
	put {\it very strong} (of order $10^{-9}$!!)
        bounds on $\alpha$ for photons and electrons, in the case
        of Scenario A only\cite{synch}.

        \item{}Present data from precision nuclear and atomic
        physics experiments puts very tight bounds on $\alpha$ for
        photons, electrons and hadrons, again in Scenario A),
        only\cite{atomic}.

        \item{}Present data from the absence of vacuum cherenkov
        effects puts interesting bounds on $\alpha$ in the case of
        Scenario A) \cite{seth}.

        \item{}Observations of bifringence effects in polarized light from
        distant galaxies puts tight
        bounds on a possible helicity dependent $\alpha$  
        \cite{helicity}.

        \item{}Observations of phase coherence in stellar and 
	galactic interferometry
        is expected, given certain assumptions\footnote{See \cite{ng} 
        for discussion of them.}, 
	to put order one bounds on $\alpha$ in the 
	near future \cite{interference}.

        \item{}Certain hypotheses about the Planck scale lead to
        the prediction of noise in gravitational wave detectors
        that may be observable at LIGO and VIRGO\cite{wave}.

        \item{}Under some cosmological scenarios, modifications of
        the form of (\ref{modified}) lead to distortions of the
        CMB spectrum that may be observable in near future
        observations\cite{CMB-distort}.

\end{enumerate}

We may summarize this situation by saying that a theory of quantum
gravity that leads to Scenario A) and predicts an energy momentum
relation (\ref{modified}) with $\alpha$ order unity is plausibly
already ruled out. This is shocking, as it was commonly said just a few
years ago that it would be impossible to test any physical
hypotheses concerning the Planck scale.

We can also mention three other kinds of experiments 
that by 2010 will have relevance for the
problem of quantum gravity

\begin{enumerate}

\item{}Evidence for or against supersymmetry may be detected at the Tev scale
in accelerators.

\item{}The equation of state of the dark energy will be measured
in near future experiments. Some proposals for dark energy are based on
modifications of energy momentum relations of the form of (\ref{modified}).

\item{}There are observations that appear to indicate that 
the fine structure constant is time
dependent\cite{fine}. These will be confirmed or go away. If the claim is
substantiated this offers a big challenge to the effective field theory
understanding of low energy physics.

\end{enumerate}

The combination of all these experimental possibilities signals that the
long period when fundamental physics developed independently of
experiment is soon coming to a close. As indicated above, the possible
experimental outcomes may rule out either string theory or loop quantum
gravity by 2010. Certain hypotheses about Planck scale physics, which lead to
preferred frame effects of scenario A) are already ruled out or tightly
constrained by observation.

\section{Postulates and main results of loop quantum gravity}

  For precision it turns out to be necessary to distinguish two forms
  of loop quantum gravity, which I will call versions I and II.

  \subsection{Postulates of loop quantum gravity I}

  What I will mean by loop quantum gravity, {\it version I} is the
theory
  which is the quantization of the Einstein's equations, coupled to
  arbitrary matter fields, in $3+1$ dimensions.

  \begin{itemize}

  \item{}{\it The quantum theory of gravity is the quantization of
general
  relativity, or some extension of it, involving matter fields, such
  as supergravity.  The quantization is done using
  standard non-perturbative Hamiltonian and path integral methods,
applied to the phase space coordinatized in terms of an
alternative set of variables. The configuration variables are
taken to be components of the spacetime connection, so that
general relativity in a certain precise sense is expressed in
terms of a gauge theory\footnote{These variables, and the
simplifications they bring about were
discovered by Sen\cite{sen} and formalized by Ashtekar\cite{abhay}
 in the Hamiltonian formalism and by Plebanksi\cite{plebanski} and
others\cite{JSS,CDJ} in the lagrangian formalism. By now several
 different connections are used in loop quantum gravity. These include the
 self-dual part of the spacetime connection\cite{sen,abhay}, and a real
 SU(2) connection introduced by Barbero\cite{barbero} and exploited by
 Thiemann\cite{thomas}. There are also alternate formulations that
 use both the left and right handed parts of the spacetime connection,
 \cite{me-holo,yime-holo}.}.}

  \item{}{\it  The quantization must be done in a manner that
  preserves the background independence of classical general
  relativity, and hence exactly realizes diffeomorphism invariance.}

  \end{itemize}

  In loop quantum gravity I the only non-dynamical structure that is
 fixed is a three manifold $\Sigma$, with a given topology and
 differential structure. There are no classical fields such as
metrics, connections or matter fields on $\Sigma$. The only exception 
is in modeling the quantization of spacetime regions with boundary,
as in the asymptotically flat or $AdS$ context, or in the presence of
a black hole or cosmological horizon. In these cases 
fields may be fixed on the boundary $\partial \Sigma$
to represent physical conditions held fixed there. 

  \subsection{The main results of loop quantum gravity I}

  \begin{enumerate}

  \item{}The states of the theory are known
precisely.
  The Hilbert space ${\cal H}^{diffeo}$ 
  of spatially diffeomorphism invariant states 
  of general relativity in
 $3+1$ dimensions has an orthonormal basis, whose elements 
 are in one to one correspondence
  with the diffeomorphism equivalence classes of embeddings of
  certain labeled graphs, called spin networks\cite{sn-roger},
  into $\Sigma$ \cite{sn1}.

  A labeled graph is a graph whose edges and vertices have attached
to them elements of a certain set of labels. In the case of pure
  general relativity with vanishing cosmological constant, the labels
  on the edges are given by ordinary $SU(2)$ spins. There are
  also labels on
  nodes of the spin networks, which are invariants or intertwiners
  from the representation theory of $SU(2)$. For details
  see \cite{sn-roger,sn1}..

  \item{}Certain spatially diffeomorphism invariant 
  observables have been constructed. After a
  suitable regularization procedure these turn out to be
  represented by {\it finite} operators on ${\cal H}^{diffeo}$, the space
  of spin network states\cite{spain,volume,volume2}.
  These include the volume of the universe,
  the area of the boundary of the universe, or of any surface defined
  by the values of matter fields.   Other operators also have
  been constructed, for example an operator that measures
  angles in the quantum geometry\cite{seth-angles}.
  These operators all preserve the diffeomorphism
  invariance of the states\cite{diffeo}.

  \item{}The area and volume operators have discrete, finite
  spectra, valued in terms of the Planck length\cite{volume,volume2}.
  There is hence a
  smallest possible volume and  a smallest possible area,
  of order of the Planck volume
and area.  The spectra have been computed in closed form.

\item{}The area and volume operators can be promoted to genuine
physical observables, by gauge fixing the time gauge so that at least
locally time is measured by a physical field\cite{me-maryland,diffeo}.  
The discrete
spectra remain for such physical observables, hence the spectra of
area and volume constitute genuine physical predictions of the quantum
theory of gravity.

\item{}Due to the existence of minimal physical volumes and areas, the
 theory has no excitations that correspond to gravitons or matter
 degrees of freedom with wavelengths shorter than the Planck
 length\cite{weaves,thomas-thesis}.

  \item{}Among the operators that have been constructed and found to
  be finite on ${\cal H}^{diffeo}$ 
  is the Hamiltonian constraint (or, as it is often called,
  the {\it Wheeler de Witt} equation\cite{ham}-\cite{thomas-ham}.)
  Not only can the Wheeler
  deWitt equation then be precisely defined, it can be solved exactly.
  Several infinite sets of solutions
  have been constructed, as certain superpositions of the spin
  network basis states, for all values of the cosmological
  constant\cite{loop1,thomas}.
  These are exact, physical states of quantum general relativity.

  If one fixes a physical
  time coordinate, in terms of the values of some physical fields,
  one can also define the Hamiltonian for evolution in that physical
  time coordinate\cite{me-maryland} and it is also given by a finite
  operator on a suitable extension of ${\cal H}^{diffeo}$ including
  matter fields. 

  \item{}The dynamics of the spin network states can be expressed
  also in a path integral formalism, called
  spin foams\cite{mike-foam}-\cite{foamreviews}\footnote{For the most 
  recent review see \cite{alejandro-review}.}.
  The histories by which spin network
  states evolve to other spin network states, called spin foam
  histories,  are explicitly known.
  A spin foam history is a labeled combinatorial
  structures, which can be described as a branched labeled two complex.
Spin foam models have
  been derived in  several different ways, and the results
  agree as to the general form of a spin foam amplitude. These
include: 1) by exponentiation of the Hamiltonian
  constraint, 2) directly from a discrete approximation
  to the classical spacetime theory, 3)  by constraining the
  summations in a finite state sum formulation of a four dimensional
  topological invariant, 4) from a matrix model on the space of
  fields over the group, 5) by postulating spacetime events are local
  moves in spin networks.

   Evolution amplitudes corresponding to the
   quantization of the Einstein equations in $3+1$ dimensions,
   are known precisely\cite{alejandro-review}
   for vanishing and non-vanishing values of the cosmological
   constant, and for both the Euclidean and Lorentzian theories.

  The sum over spin foams has two parts, a sum over graphs
  representing histories of spin networks, and, on each, a sum
  over the labels.  The sums over labels are known from both analytic
  and numerical results to be
  convergent\cite{finite-foam,finite-dan} for some spin foam
  models, including some corresponding to the quantization of
  the Einstein equations in $2+1$ and $3+1$ dimensions.

 For some spin foam model in $2+1$ dimensions, it has been shown that
 the sum over spin foam histories is Borel
  summable\cite{laurent-borel}.

  The physical inner product, which is the inner product on solutions
  to all the constraints, has an exact expression, given in terms
  of spin foam models\cite{RR-foam}.

  \item{}Matter may be added at to both the hamiltonian and spin foam
  formulations. For the hamiltonian theory it is known how to extend
  the definition of the spatially diffeomorphism invariant states to
  include all the standard kinds of matter fields, including gauge 
  fields, spinors, scalars and Kalb-Ramond fields. 
  These states are also invariant under ordinary Yang-Mills and
  Kalb-Ramond gauge transformations. 
  The forms for the
  matter field terms in the hamiltonian constraints are known 
  precisely. The spin foam models have been extended to include
  gauge and spinor degrees of freedom\footnote{To my knowledge whether 
  loop quantum gravity suffers from the fermion doubling problem is
  an open question.}.  Inclusion of matter fields does not
  affect the finiteness and discreteness of the area and volume
  observables. 
  
  \item{}Spin
  foam models appropriate for Lorentzian quantum gravity, called
  causal spin foams, have quantum analogues of all the basic features
  of general relativistic spacetimes\footnote{For more details on these
  models and the resulting physical picture, see\cite{Fotini-Wheeler}.}. 
  These include
  dynamically generated
  causal structure, light cones and a discrete analogue of
  multifingered time, which is the freedom to slice the spacetime
  many different ways into
  sequences of spatial slices\cite{F-foam}.
  The spatial slices are spin networks,
  which are quantum analogues of spatial geometries.

  \item{}Several kinds of boundaries may be incorporated in the
  theory including timelike boundaries, in the presence of both
  positive and negative cosmological constant, and null boundaries
  such as black hole and cosmological
  horizons\cite{linking}-\cite{me-holo}.
  In all these cases
  the boundary states and observables are understood in terms of
  structures derived from Chern-Simons theory.

  \item{}The boundary Hilbert spaces decompose into eigenspaces, one
  for each eigenvalue of the operator that measures the area of the
  boundary\cite{linking}.
  For each area eigenvalue, the Hilbert space is finite
  dimensional. The entropy may be computed and it agrees precisely
  with the Beckenstein-Hawking semiclassical result,
\f
S = {A[S] \over 4 \hbar G_{Newton}}
\label{bb}
\ff

 Among the boundaries that can be studied are horizons.  The boundary
 theory then provides a detailed microscopic description of the
 physics at the boundary. Furthermore, the prediction of Bekenstein
 and Hawking that an horizon should have the entropy (\ref{bb})
 is completely explained in terms of the
  statistical mechanics of the state spaces associated with the
  degrees of freedom on the horizon. This has been found
  to work for a large class of black holes, including Schwarzschild
  black holes\cite{kirill1,isolated}.

  The calculation of the entropy involves a parameter,
  which is called the Imirzi parameter. This can be understood
  either as a free parameter that labels a one dimensional family
  of spin network representations, or as the (finite) ratio of the
  bare to renormalized Newton's constant. The Imirzi parameter is
  fixed precisely by an argument invented by Dreyer, involving
  quasi normal modes of black holes\cite{dreyer}. Dreyer's argument depends
  on a remarkable precise
  coincidence between an asymptotic value
  of the quasi normal mode frequency and a number which appears in
  the loop quantum gravity description of horizons. The value of
  the asymptotic quasi normal mode frequency was at first known
  only numerically, but it has been very recently derived
  analytically by Motl\cite{motl}.  Once Dreyer's argument fixes the
  Imirzi parameter, the Bekenstein-Hawking relation (\ref{bb}) is
  predicted exactly for all black hole and cosmological
  horizons\footnote{Dreyer's calculation leads also to the conclusion that
  transitions where punctures, i.e. ends of spin networks, 
  are added or subtracted to the boundary, must be dominated by
  creation and annihilation of spin $1$ punctures.}.

  \item{}Corrections to the Bekenstein entropy have been calculated and
found to be logarithmic\cite{logcorrect}.

  \item{}Suitable approximate calculations reproduce the Hawking
  predict a discrete fine structure in the
  Hawking spectrum\cite{kirill-radiate,fineBH}. At the same time,
  the spectrum fills in and becomes continuous in the limit of
  infinite black hole mass. 
  This fine structure stands as another definitive
  physical prediction of the theory.  

  Thus, to summarize, loop quantum gravity leads to a
  detailed microscopic picture of the quantum geometry of a black hole
  or cosmological horizon\cite{isolated}. This picture reproduces
  completely and  explains the results
  on the
  thermodynamic and
  quantum properties of horizons from the work of
  Bekenstein\cite{bek1}, Hawking\cite{hawking} and Unruh\cite{unruh}.
  This picture is completely general and applies to all black hole
  and cosmological horizons.

  \item{}For the case of non-vanishing cosmological constant, of
  either sign, there is an exact physical state, called the Kodama 
  state, which is
  an exact solution to all of the quantum constraint equations, whose
  semiclassical limit exists\cite{kodama}.
  That limit describes deSitter or
  anti-deSitter spacetime.  Solutions obtained by perturbing around
  this state, in both gravitational\cite{positive}
  and matter fields\cite{chopinlee}, reproduce, at
  long wavelength, quantum field theory in curved spacetime and the
  quantum theory of long wave length, free gravitational waves on
  deSitter or anti-deSitter spacetime\footnote{For a possible
  $\Lambda=0$ analogue of the Kodama state, see \cite{mikovic}.}.

  \item{}The inverse cosmological constant turns out to be quantized
  in integral units, so that $k= 6 \pi / G \Lambda$ is an
  integer\cite{linking}.

  \item{}The thermal nature of quantum field theory in a deSitter
  spacetime is explained in terms of a periodicity in the
  configuration space of the exact quantum theory of general
  relativity\cite{positive}.

  \item{}A large class of states are known which have course grained
  descriptions which reproduce the geometry of flat
  space, or any slowly varying metric\cite{weaves,newweaves}.
  Linearizing the quantum
  theory around these states yields linearized quantum gravity, for
  gravitons with wavelength long compared to the Planck
  length\cite{graviton-weave}.  It is also understood how to
  construct coherent states which are peaked around classical
  configurations\cite{coherent}.

  \item{}A reduction of the exact physical state space to states
which
  are spatially homogeneous is known, and the reduction of the
dynamics
  to this subspace of states is known\cite{LQC}.
  (This is different from the
  usual quantum cosmology in that the reduction to homogeneous states
  is done in the Hilbert space of the full theory, rather than before
  quantization.)  The evolution of these states has been studied in
  detail and it has been found generically that the usual FRW
  cosmology is reproduced when the universe is very large in Planck
  units. At the same time the cosmological singularities are removed,
  and replaced by bounces where the universe re-expands (or
  pre-contracts).  When couplings to a scalar field are included,
  there is a natural mechanism which generates Planck scale inflation
  as well as a graceful exit from it\cite{LQC}.

  \item{}Many of these results extend to quantum supergravity for $N=1$
  and several have been studied also for $N=2$\cite{super}.

  \item{}The same methods can
  also be used to solve quantum gravity in $2+1$ dimensions\cite{2+1}
  and in
  many $1+1$ dimensional reductions of the theory\cite{1+1}.
  They also work to
  solve a large class of topological field theories\cite{tft,BF},
  giving results
  equivalent to those achieved by other methods. Further, loop
  methods applied to lattice gauge theories yields results equivalent
  to those achieved by other methods\cite{latticeloop}.
  
   \item{}In both flat space and around deSitter spacetime, one may
  extend the calculations that reproduce quantum theory for
  long wavelength gravitons and matter fields to higher energies.
 These calculations reveal the
  presence of corrections to the energy-momentum relations of the
  form of (\ref{modified}). However, now the parameters
 $\alpha$ and $\beta$ are computable constants, that depend on
  the ground state wavefunctional\cite{GP,AMU,positive}.
  These represent further predictions of the theory.

  \item{}Many of these results have been checked by being derived by
  several
  different methods, involving different regularization procedures.
  Some of these employ a high energy physics level of rigor, while
  other methods are fully rigorous, at the level of mathematical
  quantum field theory\cite{gangof5,thomas,thomas-thesis}.
  All the key results have been verified by
  being rederived with completely rigorous methods.

  \end{enumerate}

  On the basis of these results, it can be claimed that loop quantum
  gravity I is both the correct quantization of general relativity
and
  a physically plausible candidate for the quantum theory of gravity.
  It appears to provide a precise answer to the first 9 questions in my
  list.

  The failure of quantum general relativity in perturbation theory is
  explained by the fact that there are, in this quantization of
  general relativity, no degrees of freedom that correspond to
  gravitons or other perturbative quanta with wavelength shorter than
  the Planck scale.  The ultraviolet divergences are eliminated
  because a correct quantization, that exactly realizes spatial
  diffeomorphism invariance, turns out to impose an ultraviolet cutoff
  on the physical spectrum of the theory.  The assumptions
  mentioned above, that spacetime is smooth and lorentz invariant 
  at arbitrarily short
  scales, are not used in the quantization procedure, and 
  in fact turns out to
  be contradicted by the results.

  Readers with a training in perturbative quantum field theory may be 
  skeptical of these claims. There are two important things that may 
  be said in response. First the results are not about generic 
  perturbatively non-renormalizable theories. The key results of both the 
  hamiltonian and path integral quantizations follow from two necessary
  features special to gravitational theories\footnote{These are described in 
  detail in the cited references. It is fair to say that any 
  criticism of these results is ill-informed if it is not based on
  a technical understanding of how these two features are implemented in 
  the quantization procedure.}. The first is the spatial
  diffeomorphism invariance. This imposes a method of quantization
  that would fail for ordinary Poincare invariant quantum field 
  theories. This is not based on Fock space, it is based on
  a certain representation of the algebra of Wilson loop observables, 
  which allows a rigorous\cite{gangof5,thomas-thesis} formulation
  of the theory incorporating an
  exact unitary representation of the group of spatial 
  diffeomorphisms.. As a result, 
   many potential divergences are 
  eliminated by the requirement that operators be constructed by 
  regularization procedures that preserve the diffeomorphism 
  invariance of the states in the limit the regulator is removed. 
  
  The second feature is that the actions for many known gravitational
  theories can be put in the form which is closely related to a
  class of topological field theories\cite{positive,higher}. 
  These are called $BF$ theories
  because their actions are of the form of $\int Tr B \wedge F $. The 
  actions of these gravitational theories are the sum of this term with
  a constraint, non-derivative and quadratic in $B$. Theories which
  can be so expressed may be called {\it constrained topological field
  theories.}  These include general relativity in all 
  dimensions\cite{higher} and supergravity, at least in $d=4$ for
  $N=1,2$ and in $d=11$ \cite{11d}. 
  
  The combination of these two features makes possible the unexpected 
  results cited.

  It should also be said that all of the key results in the 
  hamiltonian theory, and some in the path integral theory, are
  understood completely rigorously\cite{thomas-thesis,gangof5}. A 
  reader may doubt that the world is constructed as the quantization
  of general relativity, but it is no longer an option to disagree 
  that these methods lead to a rigorously understood class of 
  diffeomorphism invariant quantum field theories in four dimensions. 
  Given the non-triviality of the existence of a class of quantum field
  theories that implement exact diffeomorphism invariance while still
  having local degrees of freedom, it is hard to believe that there are
  not important things to learn from them about how nature succeeds in
  unifying the postulates of quantum theory with the basic postulates 
  of general relativity.
  
 These claims are non-trivial, and depend on the details of the 
 construction of the hilbert space and operators involved. The point 
 is that because the construction differs significantly from that
 of a Poincare invariant local quantum field theory, the hard issues 
 are different. Ultraviolet finiteness is obtained in this case, so the
 usual worries concerning existence and consistency of the limit in 
 which the lattice spacing is removed are resolved.  One might worry 
 about taking a limit of the Planck length to zero, analogous to the
 limit of the lattice spacing to zero. But one cannot, because the
 renormalization of the Planck length is fixed to be a number of order 
 one by the requirement that the black hole entropy and the graviton 
 spectra come out right.  Furthermore, the gauge and spatial 
 diffeomorphism invariance is realized exactly, for finite $l_{PL}$,
 so there is not the usual motivation to take an ultraviolet limit to
 restore symmetries. But if the usual ultraviolet problems
 are resolved, there remain however, hard 
 issues concerning whether and how 
classical general relativity 
dominates a suitably defined low energy limit. The fact that the 
theory is well defined and finite does not, so far as we know, 
guarantee that the low energy limit is acceptable.

  Regarding these dynamical issues, at the present there are
  positive indications, but our understanding of the low energy limit 
  is far from complete. One set of issues which has been studied in a 
  lot of detail
  has to do with
  whether the action of the Hamiltonian constraint is consistent
  with a low energy limit which has  massless 
  excitations.  There is an indication that
  certain transitions, necessary for long ranged correlations
  and relativistic invariance are missing in the regularized
  hamiltonian constraint\cite{problem1,problem2}.  
  The reason appears to be that the 
  regularization procedures used involve point splitting in the spatial
  manifold $\Sigma$, but not in time. The needed terms, however, 
  are present in 
  the spin foam formalism\cite{RR-foam,F-foam}, as that is derived in 
  ways that do not depend on the $3+1$ splitting of spacetime.   
  They also appear in the hamiltonian theory for non-zero cosmological
  constant because the inclusion of $\Lambda$ imposes a quantum
  deformation on the hilbert space so that the basis elements
  are described by quantum spin networks\cite{tubes}, which 
  automatically includes the missing terms\footnote{Whether or not the
  missing terms can be derived from a regularization of the hamiltonian
  constraint for $\Lambda=0$ that involves point splitting in both 
  space and time is presently an open conjecture.}. 

  Similarly, while the problem of the recovery of general relativity
  in the low energy limit of the theory is still unsolved for zero
  cosmological constant, there is a strong indication that the
  existence of the Kodama state allows a satisfactory solution
  of the problem so long
  as the bare cosmological constant is
  non-zero\cite{kodama,chopinlee,chopin-thesis,positive}.

  \subsection{Loop quantum gravity II}

  While loop quantum gravity I so far appears to be satisfactory as both a
  quantization of general relativity and a quantum theory of gravity,
  it may very well be that the quantization of general relativity
does
  not in fact describe nature. The dimension of spacetime, physical
  degrees of freedom, and fundamental symmetries may be different from
  those which are presently observed.  It turns out that there is a
  natural class of models which generalizes loop quantum gravity
  which addresses these possibilities. These may be called
  {\it loop quantum gravity II} models\footnote{Another name for
  these theories which is sometimes used is {\it categorical state
  sum models}, because they may be formulated elegantly in the
  language of tensor categories\cite{louis-holo}.}.

To discuss these we may observe that the mathematical language of states,
  histories, boundaries and observables which is derived in the case of quantum
  general relativity can be easily generalized to give a large class
  of fully background independent quantum theories of spacetime. To
  describe the kinematics of a theory of this kind one must specify only an algebra
  (or superalgebra)
  whose representation theory is used to label the spin networks.
  The graphs on which the spin networks are based are defined combinatorially,
  so that the
  need to specify
  the topology and dimension of the spatial manifold is
eliminated\cite{F-foam,tubes}.
  In such a theory the dimension and topology are dynamical,
  and different states may exist whose coarse grained descriptions
  reveal manifolds of different dimensions and topology.

  The main postulate of loop quantum gravity II may be stated as
  follows:

  \begin{itemize}

  \item{}The states of a quantum theory of gravity are given
  by abstract spin networks associated with the representation theory of a
  given Hopf algebra or superalgebra, $\cal A$\footnote{Here a spin
  network is a graph whose edges are labeled by representations
  of $\cal A$ and whose nodes are represented by invariants of
  $\cal A$.}.

  \item{}The histories of the
  theory are given by spin foams labeled by the same representations.
  The dynamics of the theory is specified by evolution
  amplitudes assigned to the nodes of the spin foams (or equivalently
  to local moves by which the spin networks evolve).

  \end{itemize}

  Many of the results of loop quantum gravity I apply
in a suitably generalized form to loop quantum gravity II. Loop
quantum gravity
II thus specifies a large class of background independent quantum
  theories of space, time and gravitation.  There are even proposals
  that a particular form of loop quantum gravity II may be the background
  independent form of string theory\cite{mcs}.

  There are many loop quantum gravity II models that are not loop
  quantum gravity I. Examples include dynamical triangulation
  models\cite{dynamical} and causal dynamical triangulation
  models\cite{AL}-\cite{c=1}.  These take the trivial case in
  which the algebra $\cal A$ contains only the identity operator,
  but they have states which are described in terms of graphs
  and histories that satisfy the definition of a spin
  foam model. We will discuss the results achieved with these
  models below. 
  
  Finally, it should be mentioned that, at least in $2+1$ and $3+1$
  dimensions the cosmological constant is coded in a natural way in
  all loop quantum gravity theories, which is that it is related to
  the quantum deformation of the algebra of representations of the
  local lorentz group\cite{linking,2+1lambda}.
  
  One may think of loop quantum gravity II theories in the following 
  terms. Suppose we want to construct a completely background independent 
  quantum
  field theory. Such a theory must be independent of any of the
  ingredients of a classical field theory, including manifolds, 
  coordinates, metrics, connections and 
  fields. What is left of quantum theory when we remove all references
  to these structures? The answer is just algebra, representation 
  theory and combinatorics. Loop quantum gravity II models are nothing
  but a general class of quantum theories constructed using only these 
  ingredients. Consequently, one may think of generalized spin foams 
  as a kind of
  generalized Feynman diagram, in which the momentum labels are 
  replaced by representation of some algebra $\cal A$ and the energy
  and momentum conserving delta functions at the nodes are replaced by
  invariants of $\cal A$. 

  \subsection{Open questions in loop quantum gravity}

Loop quantum gravity  gives an apparently consistent microscopic
descriptions of quantum spacetime, in both Hamiltonian 
and path integral forms.  It is probably fair to say that no other
approach to quantum gravity has amassed such a long list of 
highly non-trivial results concerning quantum spacetime at the
background independent level. At the same time there remain
important open problems.  

The main open issues concern whether and how general relativity,
coupled to quantum matter fields, is recovered in a suitable low
energy limit.

For the case $\Lambda \neq 0$, there are good indications that an
acceptable solution may be achieved, based on expansions around the
Kodama state , as
described in \cite{kodama,chopinlee,chopinnew,chopin-thesis,positive}.
However, the question of whether or not the theory has a
good low energy limit
is open for general states. This includes the case $\Lambda =0$ 
in which the Kodama state does not exist.  This is a serious
problem, because it is possible that a theory may be ultraviolet 
finite, but fail to have a phase in which there is anything like
a low energy description described in terms of classical general 
relativity. This in fact occurs in some approaches to quantum
gravity such as, so far as is known, Euclidean dynamical triangulations
in $4$ dimensions\footnote{To be discussed in detail below.}  Thus,
if loop quantum gravity fails, this is likely to be how. 

To study the problem of the low energy behavior, apart from the
Kodama state,  the following research programs are underway:

 1) Renormalization group
 studies, based on a reformulation of the renormalization group to
 spin
 foam models. As a byproduct of this work it is shown that while the
 renormalization group is not a group, it does have a natural
 algebraic setting, as  a Hopf algebra\cite{f-RG,f-FRG}.

  2) The sums over labels in several spin foam models has been shown, 
  to be convergent\cite{finite-foam,finite-dan}.
  This is surprising because
  the sum over labels is analogous to the momentum integrals in
  perturbative quantum field theory.

  3) There is an understanding
  of coherent states of the quantum gravitational field, which is expected to
play a key role in understanding the low energy limit within the
Hamiltonian framework\cite{thomas-thesis,coherent}.

It should also be emphasized that the question of whether a spin foam 
model has a good
low energy limit should be asked, not just of quantum general relativity
and supergravity in
$3+1$ dimension (i.e. loop quantum gravity I), but for the whole
infinite set of theories defined by loop quantum gravity II.

There are the following possibilities:

\begin{itemize}

    \item{}A large class of loop quantum gravity theories have good
    low energy limits. In this case the existence of a low energy
    limit will be neither restrictive nor predictive.

    \item{}A restricted set, or possibly a unique, loop quantum
gravity
    theory will turn out to have a good low energy limit. In this case the
existence
    of a low energy limit will be predictive.  For example, it
    is possible that only loop quantum gravity theories
    with non-vanishing values of $\Lambda$ will have good low energy
    limits.

\end{itemize}

When a good low energy limit exists, it should be possible to discuss
perturbation theory around it.  Because studies of low energy
excitations show that there are no perturbative states around
a loop quantum gravity background with wavelength smaller than the
Planck length, perturbation theory is expected to be finite. So far, 
however, no details have been worked out beyond the linearized states. 
So this remains an important open issue.  One possible route towards 
its solution involves expanding around the Kodama state.

Another set of open issues is that of constructing Hamiltonians to get
more detailed information about the dynamics of the hamiltonian theory.
While it is important that there are many exact solutions to the 
full set of constraints, it is difficult to extract physics from most
solutions because of problems constructing fully spacetime 
diffeomorphism invariant observables. One approach that could be 
developed more is to fix a time gauge, using either boundary 
conditions or matter fields to provide the definition of a clock, and
construct the corresponding hamiltonians as operators on the space of 
spatially diffeomorphism invariant spin network states. While there
have been a few papers about implementing asymptotically flat 
boundary conditions more work needs to be done in this area as well.
Another important step would be to extend the positive
energy theorems from the classical to the quantum theory. In general 
there needs to be more development of methods to extract dynamical
predictions from the theory. 

Another key open issue is the status of global lorentz invariance.
We may note that there is no reason that quantum gravity must be
Lorentz invariant, as this is only a global symmetry of a particular
solution of the classical limit of the theory.
Global symmetries are in no way symmetries of the fundamental
theory of gravity, neither classically nor quantum mechanically. They
are symmetries of particular solutions of the classical theory.
Whether these symmetries are fully realized in the quantum states
that have semiclassical approximations corresponding to these classical
solutions is an open problem. The results mentioned suggest that the
global lorentz symmetry is not fully realized in the ordinary way.

Indeed, as mentioned,
several recent calculations indicate the presence of Planck scale
corrections to the energy-momentum relations of the
form of (\ref{modified}), effects that should be
absent were the usual lorentz transformations  exact
symmetries\cite{GP,AMU,positive}.  One issue here is that different
calculations make different assumptions about the ground state. In
some the ground state is not Lorentz invariant, hence there is no 
surprise if the perturbations around them have non-lorentz invariant
spectra. However, modified dispersion relations may also be seen 
by studying low energy exciations of a putative around state that does 
not single out a preferred 
frame\cite{positive}. The question is then dynamical: can we determine
the ground state precisely enough to discover whether the theory makes
unambiguous predictions for the parameters in the energy-momentum
relations (\ref{modified})?

  If these predictions survive further scrutiny, another important
  question is whether Scenario A) or case B) discussed in section
  4 are realized. As we discussed there, not only does each
  possibility lead to effects that are observable in present or
  near future experiments, it appears possible that in Scenario A), some
  calculational results disagree with present observations. 

  If loop quantum gravity leads to Scenario A) it may then likely be
  ruled out as a quantum theory of gravity. There is, however, a
  simple reason why we would expect case B) to be realized. This
  is that, even though there is no global lorentz invariance in
  classical general relativity, the existence of effects due
  to a preferred frame are ruled out by the condition of invariance
  under the action of the hamiltonian constraint. This is because
  in any compact region the hamiltonian constraint can generate
  changes in slicing that in any finite region are
  indistinguishable from lorentz boosts. This is true even in the
  case of solutions, like homogeneous cosmological solutions, that
  have global preferred frames.

  Now some of the key results of loop quantum gravity tell us that 
  the hamiltonian constraint can be defined and solved exactly, and
  that no anomalies are introduced into the constraint algebra by the
  quantization. This makes it very probable that any quantum
  state that is both an exact solution to the hamiltonian constraint
  and has a semiclassical limit will in that limit describe
  physics which is to leading order invariant under the action of
  the classical hamiltonian constraint. This implies the absence
  of a preferred frame of reference in the classical limit of an
  exact solution to the hamiltonian constraint.
  
  So this appears to rule out Scenario A), so long as the theory is
  defined in terms of solutions to all the constraints. 
  However there is no reason to
  expect global lorentz invariance must be
  realized as a linear rather than non-linear invariance. To the 
  contrary there is a good physical reason to expect case B), which
  is that the Planck scale can be observer independent in the limit in 
  which invariance under preferred frames is 
  realized\cite{gac-dsr,joaolee2}. 
  
  Another set of open issues have to do with the physical inner 
product.  The inner
product on diffeomorphism and gauge invariant states is known exactly
in terms of spin network states.  In Thiemann's
formalism\cite{thomas,thomas-thesis} the $SU(2)$ connection used is
real\cite{barbero} so the problem of realizing all the real
observables as hermitian operators is solved. However, the inner
product may have to be modified further to ensure that
physical states, which are solutions to all the constraints,
including the
Hamiltonian constraint, are normalizable. A complete expression for
the physical inner product is known in the spin network
formalism\cite{RR-foam}.  However, it is unlikely to have
a simple closed form.  Thus a novel feature of spin foam models is
that the physical inner product is incorporated in the path integral
that defines the physical evolution amplitudes, and it has
to be evaluated in whatever approximation scheme is being used to
pull physical amplitudes from the spin foam path integral.  Thus, 
while the solution to this problem is known in detail, it will be 
good to understand how it is implemented in detail in different 
expansions around non-perturbative states and histories.

  There are also some unresolved issues concerning the role of the four 
  dimensional diffeomorphism group in the Hamiltonian theory. This 
  comes into the details of the regularization of the Hamiltonian 
  constraint and the relationship between the hamiltonian and path 
  integral quantizations.   A set of related issues have to do with the
  relationships between the different forms of the quantum hamiltonian
  constraint arrived at by different regularization procedures and 
  different operator orderings.  We may note that the only necessary 
  condition on a candidate form of the quantum hamiltonian constraint 
  is that it have an infinite dimensional space of solutions, 
  corresponding to a theory with an infinite number of degrees of 
  freedom.  This is satisfied by Thiemann's form of the constraint, 
  and there is evidence that it is satisfied for the form of the 
  constraint which is solved by the Kodama state.  Further conditions 
  have been suggested in the past, having to do with the algebra of 
  the quantum constraints, however it seems impossible to implement 
  them in a real quantum field theory where the constraints must
  be defined as limits of regulated operators.  

Because we know that there are in fact infinite dimensional spaces of
solutions to the constraints, none of these issues 
appears to be fundamental, but they need to be resolved
nonetheless\footnote{For a novel and recent approach to related
issues, see \cite{gambini-recent}.}.

One way to summarize the status of loop quantum gravity I and II is
to state likely ways that they might fail.

\begin{itemize}
    
    \item{}Loop quantum gravity I will fail if it turns out that the 
    low energy limit of quantum general relativity coupled to matter 
    is not classical
    general relativity coupled to quantum matter fields\footnote{As mentioned 
    that there is so far no evidence for a good low energy limit for 
    zero cosmological constant, and positive, but not definitive 
    evidence for the case of positive cosmological constant.}.
    
    \item{}Loop quantum gravity II will fail if there is no 
    generalized spin foam model which has a low energy limit which
    is classical general relativity coupled to the observed standard 
    model matter fields. 

    \item{}Loop quantum gravity I or II could fail if they makes 
    predictions regarding Planck scale effects that are falsified by 
    experiment. 
    
\end{itemize}

\section{Definition and main results of string theory}

  \subsection{The definition  of a string theory}

We cannot start off the discussion of string theory with a 
list of postulates,
as we were able to do in the case of loop
quantum gravity. The reason is that many string theorists would
argue that, to the extent that string theory is the theory of nature,
its postulates have not yet been formulated. Moreover, the conceptual
ideas and mathematical language necessary to express string theory in
an axiomatic form are widely believed to remain so far undiscovered.

In this way string theory may be compared to previous research
programs such as quantum mechanics and general relativity where
several years of hard work preceded the formulation of the postulates
of the theory.  Thus, the {\it research program} called ``string theory'' 
can be taken
to be a set of activities in search of the definition of a theory
to be called ``STRING THEORY.''  What exists so far is only  a collection
of results concerning many different ``string theories.'' These
are conjectured to be each an approximate descriptions of some sector
of the so far undefined STRING THEORY.  Thus, the discovery of the 
postulates of the theory
is likely to occur close to the end of the development of the
research program, it may indeed mark its culmination.

There is of course no {\it a priori} reason to believe such a
research program will not pay off in the end.  But this situation can
complicate efforts to achieve a consensus or an objective evaluation
of the status of the theory. For this reason I propose here to
carefully separate results on the table from the exciting conjectures
that have been made. Only by doing so can we get a good idea of what 
needs to be done to
prove or disprove the main conjectures of the theory.

So my goal here will be to evaluate where string theory stands,
{\it now}, with respect to its ability to answer the questions
formulated in section 2.
To discuss the results on the table, we cannot talk about
STRING THEORY, for that does not exist as of this moment.
We must instead talk about {\it string theories},
for these are what the results in hand concern.

  Thus, by a string theory
  I will mean here what is sometimes called a perturbative
  string theory.  A more accurate name, which I will use here, is a
  background dependent string theory.  These are theories which are
  defined in terms of the embedding of two and higher dimensional
  quantum extended objects in a background classical spacetimes.
  So far as I have been able to determine,
  all of
  the firm and widely accepted
  results of string theory concern such background dependent
  theories.

  To define a background dependent
  string theory, one must first specify a classical
  background, consisting of a given manifold $\cal M$, of some
  dimension $d$ and a metric,
  $g_{ab}$. The background fields are often supplemented by certain
  other fields, which include a scalar field $\Phi$, called the
  dilaton, and generalizations of
  electric and magnetic fields, which we will denote generally
  as $A$. We then denote a choice of  background
  ${\cal B} = \{ {\cal M}, g_{ab}, \Phi, A \}$.

  There are classical theories of the motion of strings, as well as
  higher dimensional membranes in such backgrounds. 
  Examples include theories of the stretched strings
  and membranes used in musical instruments. But what makes string theory
  challenging is that not all such theories can be cast into the
  domain of quantum theory. In many cases inconsistencies appear
  when one attempts to describe a string or membrane stretched in a
  classical background in the language of quantum mechanics.

  But not always. What is remarkable is that there are some
  string theories which appear to be consistent quantum mechanically.
  They are what the subject of string theory is all about.

  Thus, the important definition to make is that of a
  {\it consistent string theory.}  This is defined as follows:

  \begin{itemize}

  \item{}A consistent
  string theory is a quantum theory of the propagation
  and interactions of
  one dimensional extended objects, closed or open, moving in a
  classical
  background, $\cal B$, which is completely consistent quantum
  mechanically. In particular it is unitary (which means quantum
  mechanics preserves the fact that probabilities always add up to
  one)
  and the energy is never negative\cite{GSW,Joe}.

 \item{}A background $\cal B$ is called consistent if one may define
a consistent perturbative string theory moving on it. Many backgrounds
  are not consistent. However there are a very long\footnote{counting
 distinct classical backgrounds as distinct.} list of
  consistent backgrounds, and some backgrounds allow more than one
  perturbative string theory to be defined on it.

  \item{}Consistent string theories are generally characterized by
  two parameters, which are a length $l_{string}$, called the
  string scale, and a dimensionless
  coupling constant $g_{string}$, called the string coupling constant.
  There may also be additional parameters
  associated with the different backgrounds. These measure aspects of
  their geometry or the values of the other background fields. In
  many cases these may be varied continuously without affecting the
  consistency of the string theory.

    \item{}A string theory is called {\it perturbative} if it describes
  interactions of strings in terms of  a power series in the dimensionless
  coupling constant $g_{string}$, such that when $g_{string}=0$
  there are no interactions.

  \end{itemize}

Now we turn to the results.

  \subsection{Basic results of perturbative string theory}

  \begin{enumerate}

  \item{}Perturbative string theories are known, which are consistent
  through second non-trivial order in string perturbation
  theory\footnote{String perturbation theory is defined as an expansion
  in the genus of the topology of the two dimensional world surface
  of the string. Leading order is a sphere, first leading order is a
  torus, etc.}\cite{GSW,Joe}.
  These include five supersymmetric string theories which are
  defined in $10$ dimensional
  Minkowski spacetime.

  The one loop consistency is well
  understood\cite{GSW,Joe}, while the two loop consistency has
  only been proved recently\cite{DP}.   Beyond two loops, there are
  partial results\cite{pasttwoloops} which support the conjecture that
  the theory is consistent to all orders.  There are intuitive
  arguments that suggest that ultraviolet divergences of the kind
  that plague conventional quantum field theory cannot occur in
  string theory.  The main reason is that the interactions of strings
  involve the breaking and joining of strings and these do not take
  place at points.  However, a string theory can fail to be
  consistent for other reasons.  There may be infrared divergences, or
  ambiguities in the definition of the amplitudes, there can be
  anomalies in the action of the lorentz boosts,  or the theory may
  fail to be unitary.  The problem of the
  consistency of perturbative string theory appears to be very
  challenging, and the proof of consistency at two loops required a
  very impressive technical {\it tour de force\cite{DP}.}

  As it does not appear to be widely appreciated that the consistency
  of string perturbation theory is still open\footnote{Among other 
  claims that apparently did not stand up, a paper of
  Mandelstam\cite{mandelstam} is sometimes cited. However, the 
  opinion of experts familar with the technical issues involved
  appears to be that this paper does not contain a
  satisfactory or complete  proof of finiteness and 
  uniqueness of superstring amplitudes to all orders. For a discussion 
  of some of the technical issues involved, see \cite{marshakov} and
  \cite{kaku}, who says in part, (p 226) 
{\it `` Being able to isolate the singularities... is a great
step towards solving the main problem facing string perturbation 
theory, i.e. rigorously proving finiteness to all orders... However, 
there are many delicate points related to the question of the 
cancelation of these diverges that have not been totally solved.
Although preliminary results are encouraging, the rigorous proof of the
cancelation of divergencees is still an outstanding problem''}.}, 
I quote here from
  a recent paper by experts in the field, which announced the proof of
  consistency at the two loop level.

  \begin{quote}
      Despite great advances in superstring theory, multiloop amplitudes
      are still unavailable,
almost twenty years after the derivation of the one-loop amplitudes by
Green and Schwarz
for Type II strings and by Gross et al. for heterotic strings.
The main obstacle is
the presence of supermoduli for worldsheets of non-trivial topology. 
Considerable
efforts had been made by many authors in order to overcome this 
obstacle, and a chaotic
situation ensued, with many competing prescriptions proposed in 
the literature. These
prescriptions drew from a variety of fundamental principles such 
as BRST invariance and
the picture-changing formalism, descent equations and Cech cohomology, modular
invariance, the light-cone gauge, the global geometry of the Teichmueller curve,
the unitary gauge, the operator formalism, group theoretic methods,
factorization, and algebraic supergeometry. However, the basic problem was that
gauge-fixing required a local gauge slice, and the prescriptions ended 
up depending on the
choice of such slices, violating gauge invariance. At the most 
pessimistic end, this raised
the undesirable possibility that superstring amplitudes could be ambiguous, 
and that it
may be necessary to consider other options, such as the Fischler-Susskind
mechanism\cite{DP}.
  \end{quote}

  This situation is  a bit disappointing, given that the main claim
  for string theory as a quantum theory of gravity is that it alone
  gives a consistent perturbation theory containing gravitons. After
  all, supergravity theories, which are ordinary field theories which
  extend general relativity to incorporate supersymmetry, are also
  consistent in perturbation theory at least to the two loop
  level and $N=8$ supergravity in four dimensions is expected
  to be consistent at least to five loops\cite{sugra2loop}.  The difference is
  that there are reasons
  to expect that supergravity theories become inconsistent at
  some point beyond two loops,
  while no reason is known that the technical
  difficulties that have blocked a proof of the consistency of 
  perturbative string theory
  cannot someday be overcome.

  It is further known that bosonic string perturbation theory is not Borel
resumable and this is conjectured to extend to superstring 
theory\cite{notborel}. This means that the theory cannot be
  defined completely
  by perturbation theory because there may be excitations of the full
  theory that are not captured in the perturbation theory.
  
  From now on, when I label a string theory ``consistent'' I really 
  mean that it is known to be consistent to one or two loop order, 
  and that no reason is known why the conjecture of all orders 
  consistency may not apply to it.

  \item{}There are consistent string theories with spatially closed
  boundary conditions (so the string is a closed loop)
  and with open boundary conditions.
  The latter, called
  open string theories have spectra that include
  the quanta of Yang Mills fields\cite{GSW,Joe}. The
  Yang Mills coupling $g_{YM}$
  is related to the string coupling by $g_{YM}^2 =
  g_{string}$\footnote{In four dimensions where the Yang-Mills coupling
  is dimensionless.}.

 \item{}Consistent closed string theories have spectra that include
 massless gravitons propagating on the background\cite{GSW,Joe}.
 They couple with a
 Newton's constant given by\footnote{Again, in $d=4$.}
 $G_{Newton}= g_{string} l_{string}^2$.

 \item{}All string theories which are known to be consistent to
 second order are
supersymmetric. There are some non-supersymmetric string theories
which appear to be consistent at least to first non-trivial order.
While not supersymmetric, these  have spectra in which
fermions and bosons are grouped in pairs of equal mass\cite{nonsuper}.
As this is not a feature of our world, if
string theory is true it must be that supersymmetry (or at least
fermi-boson matching) is spontaneously broken.

 \item{}Many known consistent string theories have backgrounds
 that are $10$ dimensional manifolds, or  can be understood
 as arising from $10$ dimensional manifolds by compactifications and
 identifications\cite{GSW,Joe}.
 A simple class of example of such consistent compactified
 backgrounds are those cases in which
 the compactified $d$ dimensional manifold is a $d$-torus.

 \item{}A necessary\footnote{There is a variant of string theory,
 called non-critical string theory \cite{NCS}, in which the conformal
 anomaly does not vanish perturbatively, but it is claimed, under
 certain conditions, it may vanish as a consequence of a 
 non-perturbative regularization scheme. The literature on this
 subject describes some interesting ideas, however I do not understand
 the status of the claims made well enough to include them in this
 review. } condition for a
 perturbative string theories to
 be consistent is that the two dimensional world sheet quantum
 field theory that defines the theory be
 conformally invariant\cite{GSW,Joe,CFTstrings}. This
 means that the  conformal anomaly on the two dimensional
 worldsheet vanishes.  To leading order in $l_{string}$ this
 condition is equivalent to the Einstein equations of the
 background manifold\footnote{Corrections to the background
 field equations have been worked out.}.

 \item{}There are a very large number, perhaps infinite, of
 backgrounds of the form of four dimensional flat spacetime producted
 with a six dimensional compact manifold.  A large class exists where
 the six dimensional compact manifold is a Calabi-Yau manifold. There
 are estimated to be on the order of at least $10^{5}$ distinct
 such manifolds.

 \item{}Most of the known backgrounds of the form of
 $Mink^{10-d} \times Compact^{d}$ have
parameters  that measure the
 geometry of the compactified manifold, in many cases these parameters
 can vary over the $10-d$ dimensional manifold and hence become scalar
 fields, called moduli fields, on the $10-d$ dimensional manifold.
In many cases the energy does not depend on the values of the moduli
parameters, so the fields that represent them on
$Mink^{10-d}$ are massless.

No consistent string theories are known that have no massless
scalar fields. 

\item{}These include the case $10-d=4$, which is so far at least
supported by all observations, but consistent backgrounds
exist for any $d$ up to $9$.

\item{}There are transitions in string theory in which the
topology of the $d$ dimensional compact manifold
changes\cite{phasechange}.

\item{}There are consistent string backgrounds for $4$ large,
uncompactified dimensions that correspond to a large range of
possible values for the number of generations,  for
the number of Higgs fields and for the gauge group. Thus, string theory
makes no prediction for these characteristics of the standard
model\cite{varietyofstrings}. (See figure 1.)

 \item{}There are so far not known any consistent, stable, 
 string backgrounds
 of the form of DeSitter spacetimes times a compact
 manifold\cite{stringcosmoproblem,andyds}\footnote{The reason is
 related to the fact that there is no unitary representation of the
 supersymmetric extension of the symmetry group of deSitter spacetime.
 There have been intriguing suggestions from string theorists
 about
 quantum gravity with a positive cosmological constant, a few of them
 involve string theory explicitly. However Ed Witten recently wrote,
``In fact, classical or not, I don't know any clear-cut way
to get de Sitter space from
string theory or M-theory. This last statement is not very surprising
given the classical no go theorem. For, in view of the usual problems
in stabilizing moduli, it is hard to
get de Sitter space in a reliable fashion at the quantum level given
that it does not arise
classically\cite{stringcosmoproblem}.''}.

\item{}More generally all known consistent, stable string theories
have time-like or null 
killing fields. No consistent stable string backgrounds 
are known which are time dependent\footnote{This limitation is tied to 
supersymmetry, because the closure of even the $N=1$ supersymmetry
algebra contains the hamiltonian, and that is only well defined on
backgrounds that possess a timelike or null killing field.}. 

 \item{}There are an infinite number of consistent backgrounds that
include
 structures known as $D$-branes\cite{Joe}.
 These are submanifolds of various
 dimensions embedded in the background on which open strings may end.
 These branes may be charged, with respect to some of the generalized
 electric and magnetic fields. The $D$-branes have dynamics induced
by their coupling to the strings and charges. This includes Yang-Mills
 fields which propagate on the branes\cite{joe-branes,branes-review}.

  \item{}By a careful choice of arrangements of several branes,
 intersecting at carefully chosen angles, one can construct a string
 theory background whose low energy limit has some features of the
 supersymmetric
 standard model, including chiral fermions and parity violating gauge
 couplings\cite{mostreal}.  However, many other consistent backgrounds
 exist which
 do not have these features\cite{varietyofstrings}.

 \end{enumerate}

 \begin{figure}
\center{
\epsfig{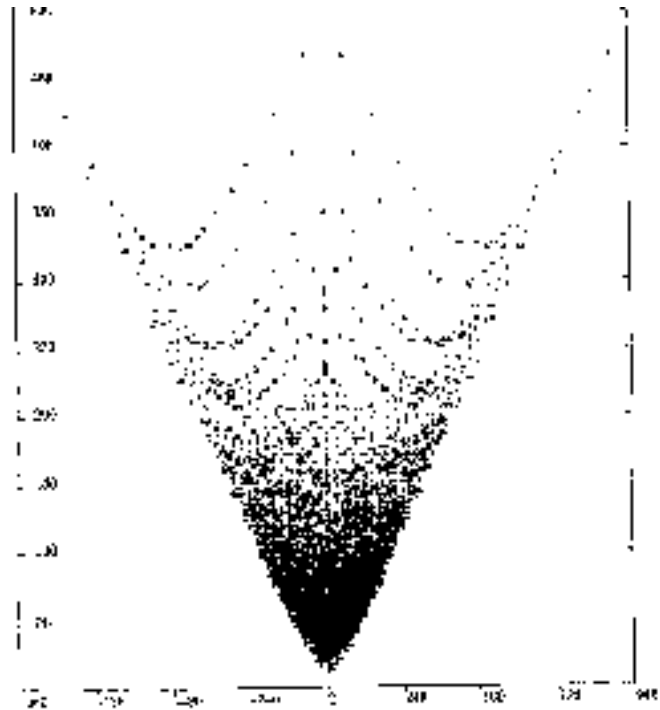}}
\caption{A sample of the phenomenology predicted by different
     consistent string theories. The vertical axis is the number of
     Higgs fields, up to 480, the horizontal axis is related to the number of
     left handed fermion fields minus the number of
     right handed fermion fields.
     According to string theory we could equally well live in any
     of these universes. From
     \cite{varietyofstrings}. }
\label{many}
\end{figure}

 \subsection{Results and conjectures concerning black holes}

 String theory also has led to results which
 are relevant for the understanding of black holes. To express them one
 has to know that in the state space of a supersymmetric theory there
 is a subspace in which a fraction of the supersymmetry
transformations
 are broken, leaving still unbroken a number of supersymmetries at
least twice
 the dimension of the spinors in that dimension. These are known as
 $BPS$ states and they have special properties\cite{BPS}. In particular,
aspects
 of the spectra of the Hamiltonian are strongly constrained by the
 supersymmetry algebra. One way to characterize $BPS$ states is that
 they are states in which certain of the generalized electric and
 magnetic charges are, in appropriate units, equal to their masses.

 Classical supergravity has $BPS$ states (i.e. classical solutions),
 among which are black holes
 whose charges are equal to their masses\cite{BPS}.  These are also called
 extremal because there is a theorem that the charges cannot exceed
 their masses.  This indeed suggests a relationship between
 supersymmetry and the properties of black holes.  The extremal
 black  holes have zero
 Hawking temperature  but nonzero Bekenstein entropy\cite{BPS}.

The results in string theory do not concern, precisely, black holes, as they are
found in a limit in which the gravitational constant is turned off.
But they concern
systems with the same quantum numbers as certain black holes, which,
it may be argued, may become black holes if the gravitational constant
is turned up to a sufficiently strong value.  Still they are very
impressive,

 \begin{enumerate}

 \item{}For certain compactifications, with $d=3,4$ or $5$ flat directions,
 and in the limit of vanishing $g_{string}$, and hence
 $G_{Newton}$,
 there are $BPS$ states of string theory including $D$-branes,
 which have the same mass,
 charges and angular momenta of an extremal black hole in $d$
 dimensions. The number of such states is in all cases exactly equal
to the
 exponential of the Bekenstein entropy of the corresponding black
hole\cite{SV,otherbh,nearextremal}.

 \item{}If one perturbs away from the $BPS$ condition for the string
 theory states, to a near extremal condition, and constructs a
thermal
 ensemble, the spectrum of the Hawking radiation from the
 corresponding near extremal black hole is reproduced exactly,
 including the grey body factors\cite{nearextremal}.

 \end{enumerate}

 These results are very impressive; the agreement between the formulas
 obtained for entropy and spectra between the $D$-brane systems and
 black holes are staggeringly precise. It is hard to believe that this
 level of agreement is not significant. At the same time, there are
 two big issues. First the $D$-brane systems are not black holes.
 Second it has not been found possible to extend the results away
 from the neighborhood of extremal, $BPS$ states, so as to apply
 to ordinary black holes.

 We are then left with a conjecture:

      \begin{itemize}

       \item{}{\bf Black hole conjecture.}
       If one turns the gravitational constant up in the
       presence of a thermal ensemble of states which as described
       above, reproduce the entropy and temperature of an extreme or
       near extremal black hole, one can construct a string theoretic
       description of quantum black hole spacetimes.  This will
       extend also to far from extremal black holes.

  \end{itemize}

  We may note that some arguments, non-rigorous but physically
  motivated nevertheless, support the conjecture that the states
  in string theory which correspond to a Schwarzschild black hole
  have an entropy proportional to the square of the
  mass\cite{stretched,matrixbh}.
  However the constant of proportionality is so far not
  predicted.

 \subsection{Results and conjectures concerning dualities}

 In section 3 I emphasized the importance of the notion of duality in
 both
 string theory and loop quantum gravity. There are indeed a
 number of very interesting results concerned with how duality is
 realized in string theory.  These motivate a number of conjectures
 which, if true, are quite important for the physical interpretation
 of string theory. As some of the conjectures remain unproven,
 however, it is important  to distinguish
 results from conjectures.

 \subsubsection{T duality} In compactifications on tori, or more
 generally where the compact manifold has non-trivial $\pi^1$, the
 string theory spectrum has states distinguished by a winding number
 around a circle as well as the usual vibrational modes. In all these
 cases there is a symmetry in which one exchanges winding and
 vibrational modes and, in units of the string scale, $l_{string}$,
 takes the radius  of the circle, $R$ to $l_{string}^2/R$.

 $T$-duality appears to be a general property of string
 theories\cite{GSW,Joe}.
 It depends neither on supersymmetry nor on criticality and so
 appears to be true for all string theories.
 However,
 the next two cases require some care as there are some unproven
 conjectures.

 \subsubsection{$S$ duality}

 $S$ duality is inspired by the old observation, which goes
 back at least to Dirac, that electromagnetism
 is almost invariant under an exchange of electric and
 magnetic fields.
 The idea is that if we had a theory with magnetic monopoles, with
 magnetic charge $g$ as well as ordinary
 particles with ordinary electric charge $e$,
 then the theory of Maxwell, modified to include the magnetic
 monopoles, appears symmetric under an exchange
 $e \leftrightarrow 2\pi /g$. Thus, if
 \f
 g= 2\pi /e
 \label{mono}
 \ff
 the theory
 might be symmetric under the symmetry operation in which
 $e \rightarrow 1/e$ and electric charges and magnetic monopoles
 are exchanged.

 In certain supersymmetric Yang-Mills theories, this appears to be
 the case, at least to some approximation\cite{S-sym}. This is
 because the theory
 has solitons which are magnetic monopoles which satisfy
 eq. (\ref{mono}). There is one theory, the $N=4$
 supersymmetric Yang-Mills theory in $4$ dimensions in which
 there is good evidence for this at least in the $BPS$ sector
 of the theory\footnote{Very recently there are results
 which strongly support the conjecture that
 supersymmetric Yang-Mills theories are $S$-dual\cite{sdualmatrix}
 to all orders in perturbation theory.}.

 In string theory there are several related results.

 \begin{itemize}

 \item{}For many string backgrounds $\cal B$ there
is an $S$-dual background ${\cal B}^\prime$ such that the free string
 spectra and $BPS$ subspace on $\cal B$ can be mapped onto the free
 spectra and $BPS$ subspace on ${\cal B}^\prime$ with $g_{string}$
 taken to $1/g_{string}$\cite{S-string}.

 \item{}In some cases $\cal B$ and ${\cal B}^{\prime}$ are the
 same, and the duality maps the $BPS$ sector of a single theory
 to itself. In this case we may speak of the theory being
 self-dual, at least on the $BPS$ sector.

 \end{itemize}

 This is quite an impressive fact, as it tells us that there
 are indeed theories, in  which a generalization of
 electro-magnetic duality holds exactly, at least in a sector
 of the state space.  It is then very interesting
 to ask whether the duality transformations hold exactly in string 
 theory
 only on the $BPS$ sectors of the theory, or extend to the full
 theories in question.

The answer depends on whether the existence of the duality is
a consequence of the $BPS$ conditions, or is an expression of a deeper
property of the dynamics of string theory.  It is true  that the
 supersymmetry algebra strongly constrains the spectra and
 degeneracies of the $BPS$ sector, because the Hamiltonian is
 part of an algebra that generates the spectrum.  If this is all
 there is to it, it is an impressive result, but it would not be
 expected that the duality would apply to the whole theory.

 However many string theorists believe that $S$ duality is a general
 property of string theories, This conjecture may be stated as

 \begin{itemize}

     \item{}{\bf $S$ duality conjecture.} Whenever an $S$ duality
     exists in the $BPS$ sector of a string theory  
     it extends to an isomorphism on the
     full Hilbert spaces of the theories in question.

\end{itemize}

There are a few results in string theory concerning the spectra of
non-$BPS$ states, and they do show that a duality transformation
continues to exist, at least approximately, to leading order in
departures from the $BPS$ condition\cite{non-BPS}.
Were this not the case the
$S$ duality conjecture would be falsified. At the same time, to my
knowledge, there is no stronger result and no proof supporting
the $S$ duality conjecture in  string theory.

 \subsubsection{String/gauge theory dualities}

 These are a new kind of duality which connects, not different
 string theories, but string theories and gauge theories.
 They are very reminiscent of the original ideas of duality.
 However this idea has been realized in the last few
 years in a novel way, in which a string theory in $d$ non-compact
 spacetime dimensions
 is related to a gauge theory in $d-1$ spacetime dimensions.

 The reason for the difference in dimensions can be explained by
 a remarkable argument, due originally to Polyakov\cite{polyakov-extra}.
 He observed
 that a string theory may be expressed as a two dimensional
 quantum field theory on the two dimensional worldsheet of the
 string. Among the fields that live on the worldsheet
 are the imbedding coordinates,
 $X^a (\sigma)$, where $X^{a}$ are $d$ coordinates in the
 $d$-dimensional background
 spacetime and $\sigma$ are the two dimensional coordinates
 on the worldsheet.  However, Polyakov noted that to make the
 worldsheet theory one needs also the metric on the worldsheet
 $h_{\alpha \beta}$, so as to form the action,
 \f
 I=\int d^{2}\sigma \sqrt{h}h^{\alpha \beta}(\partial_{\alpha}X^{a})(
 \partial_{\beta}X^{b} ) g_{ab} (X(\sigma))
 \ff
 Here $g_{ab}$ is the metric of the background spacetime.

 The two dimensional coordinates of the worldsheet can be fixed so that
 \f
 h_{\alpha \beta}= \eta_{\alpha \beta} e^{\phi}.
 \label{conforalg}
 \ff
 where $\eta$ is the metric of flat, two dimensional spacetime.
 This leaves
 unfixed the third component of the metric, represented by $\phi$,
 which we see here is the conformal factor.

 Polyakov noticed that the quantization of the $X^{a}$ fields on
 the worldsheet will in general give rise to a  conformal
 anomaly. This will occur in spite of the fact that the classical
 action is conformally invariant. This gives rise to a dynamics for
 the $\phi$ field of the form
 \f
I^{\prime}=\hbar \int d^{2}\sigma \sqrt{h}h^{\alpha \beta}
(\partial_{\alpha}\phi ) (\partial_{\beta}\phi)  \ldots
 \ff
 Thus, if this conformal anomaly is not cancelled it is as if
 the string moves in an $d+1$ dimensional spacetime background,
 whose coordinates are $X^a, \phi$.

 Now for what are called critical string theories, the anomaly is
 cancelled by factors coming from the ghosts, which must be there in
 turn because of the gauge fixing down to the gauge (\ref{conforalg}).
 The resulting string theories have massless degrees of freedom.
 However, for some gauge theories, like $QCD$ we do not expect
 there to be massless gluons in their spectra, due to confinement
 or, in some cases, to a Higgs effect. For such theories there
 cannot correspond a critical string theory.  However, there
 are still general arguments, based on the idea of duality,
 that suggest that some kind of string theory should be related
 to any gauge theory. The resolution of this puzzle is that for
 such gauge theories there should correspond a non-critical
 string theory. But then Polyakov's observation suggests that
 the dual string theory should then appear to live in a spacetime
 of one additional dimension.

 Moreover it is not hard to see that if the background spacetime
 is a Minkowski spacetime, the spacetime of one higher dimension
 that is created is an anti-deSitter (AdS) spacetime. The original
 Minkowski spacetime can be thought of as part of the boundary of that
 Anti-deSitter spacetime\cite{polyakov-extra}.

Thus, the suggestion is that to any non-Abelian gauge theory
that does not have massless gluons in its spectra in $d$
dimensional Minkowski spacetime there should correspond a
string theory in a $d+1$ dimensional anti-deSitter spacetime.

There is a further argument that suggests a relationship between
quantum field theories on $Minkowski^{d}$ and on $AdS^{d+1}$. This is
that the symmetry group of $AdS^{d+1}$ spacetime is the same as the
conformal group on $Minkowski^{d}$.  This suggest the relationship
between a gauge and a string theory on these two spacetimes
should be especially tight in any case in which the
gauge theory on $Mink^{d}$ is conformally invariant.

Now {\it classical}
gauge theories in $d=4$ are conformally invariant so long
as they don't couple to massive fields.  But in general the
conformal invariance is broken by quantum corrections. However
there are a few cases in which supersymmetric gauge theories
are known, in perturbation theory, to have vanishing $\beta$
functions, which means that they are, in perturbation theory,
exactly conformally invariant. One of these is the most supersymmetric
non-gravitational theory that exists in four spacetime dimensions,
the so called ${\cal N}=4$ supersymmetric Yang-Mills theory.
It is then natural to guess that in such cases there could
be found interesting results connecting them to string theories.

Arguments and conjectures concerning a possible connection
between string theory and supersymmetric Yang-Mills theory
were put forward in 1997, first by
Maldacena\cite{juan-conj}, Witten\cite{witten-CI}, and
 Gubser,  Klebanov and Polyakov\cite{GKP}. Since then a large number of
 very interesting results have been found in this direction.
 While it is without doubt that these results are highly significant,
 this is also an area in which it is necessary to distinguish
 results from conjectures.  Let us start with the
 results\footnote{The argument of this section was developed in
 collaboration with Matthias Arnsdorf. A more detailed
 version of this argument is presented in \cite{matthiaslee}.},

 \begin{itemize}

 \item{}{\bf Conformal induction\cite{witten-CI,matthiaslee}.}
 Consider a quantum theory, $T_1$ defined
 by a path integral, on a background $\cal B$ whose spacetime is of
 the form of ${\cal M}^{d} $ where  ${\cal M}^{d}$ is either 
 an anti-deSitter spacetime, or a
more general spacetime that is asymptotically anti-deSitter. Then it can
 be shown that this spacetime has a boundary whose timelike component
 is isomorphic to
 $M^{d-1}$ where $M^{d-1}$ is flat spacetime. One
can then argue generally that one can define a quantum field theory on
 the boundary by evaluating expectation values of local operators in
 the theory $T_1$ in which all the operators are taken to the
boundary.
 One can also argue that this new theory must be conformally invariant
 (or more specifically have exact scale invariance, with perhaps spontaneously
 broken conformal invariance.)

 Furthermore the same holds for quantum theories defined on spacetimes
 of the form given, producted by a compact manifold. 
 
 Thus in these cases a conformal field theory on the boundary is
 defined from a subset of the observables of the bulk theory, those
 in which N point functions for fields are evaluated all on the
 boundary.  We may say that the boundary theory is {\it conformally induced}
 from the bulk theory. In such cases, some
 of the observables of the bulk theory are computable in terms of
 $N$-point functions of the boundary theory. But in the general case
 there is no reason to believe that the two theories are isomorphic,
 for there will generally be observables of the bulk theory that are
 not computable in terms only of $N$-point functions on the conformal
 boundary.  One reason is that there are components of the boundary
 of an AdS spacetime besides the timelike component which is
 isomorphic to $M^{d-1}$. This includes future and past timelike infinity.

 \item{}There are many results that suggest that the
 conformal induction of linearized supergravity on
 $AdS^5 \times S^{5}$ is a certain limit of the
 $N=4$ supersymmetric Yang-Mills theory on four dimensional flat
 spacetime\cite{witten-CI,AdS-review}.

 \item{}There are additional results that strongly indicate that the
 conformal induction of supergravity on various asymptotically
 $AdS^5$ spacetimes is
related in this way to $N=4$ super-Yang-Mills theory by reduction
 or symmetry breaking\cite{AdS-review}.
 Among them are results that are consistent
 with the conjecture that the linearization of supergravity
 on the AdS-Schwarzschild solution
 induces a thermal state in the Yang-Mills theory\cite{AdS-thermal}.

 \item{}Still other results of this kind hold for other $d$.

 \item{}It is not, however, completely established whether perturbative string
 theory is well defined on an $AdS^{5}\times S^{5}$ background, beyond
 the supergravity approximation.
 While the classical action for a
 free string on the $AdS^{5}\times S^{5}$ background has been
 constructed\cite{arkady}, there is no gauge in which it is a free theory. 
 Thus the
 free string theory cannot be solved exactly on an $AdS^{5}\times
 S^{5}$ background, as it can be in the case of flat ten dimensional spacetime.
 Instead, to define the free string theory one has to treat it as an
 interacting two dimensional quantum field theory, defined on
 the string world sheet\cite{AdS-problem}.

 The resulting theory has been studied and, in some particular
 examples, shown to be a conformal field theory at least the
 one loop level\cite{ADN}.  There are also arguments that the theory
 remains a conformal field theory to all orders\cite{renata-arivand,ADN}.
 These results are very reassuring, but we are still apparently lacking
 a general proof of the consistency of the interacting string theory,
 as is possible on flat space through at least the two loop order.

 \item{}A certain limit of $AdS^5 \times S^5$ is known which is
 a plane wave\cite{PPwave}.
 This is gotten by expanding the metric around a null
 geodesic that circles the sphere, while remaining at the center of
 the $AdS^5$.   In this case the free quantum
 string theory can be constructed and solved explicitly.
  The analogous
 limit can be constructed in the $N=4$ super Yang-Mills theory.
 The resulting spectra matches
 that of the string theory on the plane wave.

 \item{}There is a general result of axiomatic quantum
 field\cite{rehren}
 theory that, given an exactly conformally invariant
 quantum field theory, without anomaly, on $Minkowski^{d}$,
 constructs an axiomatic quantum field on $AdS^{d+1}$. This
 result is rigorous, but it requires that the theory on
 $Minkowski^{d}$ have no anomaly in any of the generators of the
 conformal group. However, the supersymmetric Yang-Mills theory
 do not satisfy this, at least in perturbation theory\cite{anti-rehren}.
 While they
have $\beta=0$, and hence no anomaly in the action of  dilitations,
 they have anomalies in the action of the large conformal transformations.
 Hence it appears that the gauge theory cases studied are not
 examples of this particular version of the correspondence.

  \end{itemize}

  As in the case of $S$ duality there are conjectures which many
  string theorists believe which, if true, greatly extend
  these results.

  \begin{itemize}

   \item{}{\bf Maldacena conjecture.}
   There is an isomorphism between $N=4$ supersymmetric
      Yang-Mills theory in $d=4$ and ``string theory on $AdS^5 \times
      S^5$."  This is sometimes called the
      Maldacena conjecture\cite{juan-conj}.

      There are actually a number of different conjectures that are
      often conflated in discussions. The Maldacena conjecture, as I
      have stated it, is the strongest of them\footnote{For more
      details on the different possible versions of the Maldacena
      conjecture, see \cite{matthiaslee}.}.

      Even stated as such, one might mean two different things by the
      Maldacena conjecture.

      \item{}{\bf Maldacena I.}  String theory ``on an $AdS^{5} \times S^{5}$
      background'' can be given a precise, consistent  non-perturbative
      definition, as can $N=4 $
      supersymmetric Yang-Mills theory in 4 dimensions.  After they
      are both defined, each independently of the other,
      it can be shown that they are isomorphic.

      \item{}{\bf Maldacena II.}  String theory  ``on an $AdS^{5} \times S^{5}$
      background'' can be given a precise, consistent  non-perturbative
      definition {\it by assuming a certain correspondence between it and
      $N=4$ supersymmetric Yang-Mills theory.}

      \end{itemize}

      The first thing that must be said in evaluating the evidence for
      either of these conjectures, is that they are each logically
      much stronger than the conjecture of conformal induction, as I
      stated it above.   In fact conformal induction is a very general
      property, and it may be supported by general arguments
      that assume only the existence of a bulk and boundary theory on
      an $AdS$ or asymptotically $AdS$ spacetime. These arguments do
      not assume any special properties of supersymmetric theories or
      gravitational theories and, indeed, there are known examples
      where such a correspondence holds for non-gravitational and
      non-supersymmetric theories.

      There is unfortunately a lot of confusion about this in the
      community, due perhaps to the fact that the papers of Witten
      and Klebanov et al that followed Maldacena conjecture were
      mainly concerned with presenting arguments for the weaker
      conjecture of conformal induction. As a result, many members of
      the string theory community appear not to have noticed, or not
      to be concerned with the fact, that the conjectures are very
      different, and have rather different implications.

      Many results have been obtained supporting some version of
      a correspondence between quantum theories in AdS spacetimes and
      gauge or conformal field theories on their
      boundaries. Given the existence of
      different conjectures, care is required in evaluating which
      conjectures are supported by which results.

      When carrying out this analysis, it is important to remember a
      basic point of logic. In a case in which two
      conjectures are each consistent with a given set of results, and one
      conjecture is logically stronger than the other, it follows that the
      evidence may be taken to support only the  weaker conjecture.
      Only results which are not consequences of the weaker conjecture
       may be taken as evidence for the stronger
       conjecture\footnote{Some readers have questioned this principle of
       logic, so let me give a simpler application of it.
       So far the 
       friends I had in high school are all, to my knowledge still alive. 
       There are two possible
       conjectures that would account for this observed fact.  The first
       is that we are still much younger than our common
       statistical life expectancy.  The second is
       that we are all immortal.  The second is stronger than the
       first, because it logically implies it.  No amount of wishful
       thinking
       will convert what is decent evidence for the weaker
       conjecture into evidence for the stronger conjecture.}.

      In fact, almost all of the results found concerning an
      AdS/CFT correspondence are explained by the weaker conjecture
      of conformal induction\footnote{For results which appear to
      be consistent with conformal induction but are possibly inconsistent with
      stronger conjectures, in that they seem to indicate that 
      different bulk spacetimes which agree asymptotically may not be
      distinguished by the boundary conformal field theory, see \cite{viqar-ads}.}.  
      For example, all of the results
      concerning matching $N$ point functions between classical
      supergravity and the quantum Yang-Mills theory are of this
      kind, as are the results concerning matching (up to overall
      constants) of entropies for thermal states\cite{matthiaslee}.

      In discussing this situation one may ask the following question.
      Suppose that it turns out
      that there is no consistent {\it interacting} string theory
      on an $AdS^{5} \times S^{5}$ background\footnote{Not that there
      is reason to think this is the case, but just to make the
      following argument.}.  Would this contradict
      any of the results so far found which are used as evidence for
      an $AdS/CFT$ correspondence?

      If the answer is no then that
      evidence is completely consistent with the possibility that the
      strongest general result which holds between the gauge theory
      and the gravity theory is a form of conformal induction.
      Further,  as
      no interacting string theory may in this case exist, it holds only between
      either the free string theory, or supergravity,
      expanded on the $AdS^{5} \times S^{5}$ background
      and a certain limit of the $N=4$ supersymmetric Yang-Mills theory.

      Of interest in this regard is the recent work connecting
      the plane wave limit of $AdS^{5}\times S^{5}$ with
      a similar limit in the supersymmetric Yang-Mills
      theory\cite{PPwave}.
      This work is extremely interesting and introduces novel techniques which
      illustrate how a string theory may arise from a gauge theory.
      It is then of great interest to establish which version of
      an $AdS/CFT$ connection is supported by these results.
      The key question is whether these results are implied
      by a combination of conformal induction and the $BPS$ conditions.
      If they are then, as interesting as they are, these do not
      support the Maldacena conjecture over conformal induction.
      One might think that conformal induction is not involved, as
      the limit taken involves expanding the spacetime around a
      trajectory far from the boundary. However, the $BPS$ conditions
      do play a role in the derivation, so it is possible
      that the agreement found is simply a consequence of the fact
      that the same supersymmetry algebra acts in the plane wave spacetime
      as on the related limit of the boundary gauge theory, and the
      correspondence found is a consequence only of the fact that the
      extended supersymmetry algebra constrains some feature of the
      spectra of the theories.
      Results which settle this issue would clearly be of interest.

      In any case, apart perhaps from the Penrose limit, there appear
      to be so far no results
      relevant for interacting strings on the $AdS^{5}\times S^{5}$
      background.   This means that it is not yet possible to claim
      evidence  any conjecture stronger than conformal induction,
      except perhaps in the Penrose limit.

      A final point that must be made concerns the status of
      $N=4$ supersymmetric Yang-Mills theory. That theory is well
      defined perturbatively, and it is known to have a vanishing
      $\beta$ function.  There is also evidence that it is exactly
      $S$-dual to all orders in perturbation theory.
      
      This suggests two beautiful conjectures.
      First that this theory is well
      defined also non-perturbatively.  Second, that it is an exactly scale
      invariant quantum field theory.  However, as of this date, the
      theory has not been given a non-perturbative definition.  The
      usual route to a non-perturbative definition of a gauge theory
      is through a non-perturbative regularization such as the lattice
      theory.  Given the fact that the $\beta$ function vanishes, it
      is to be expected that {\it were such a non-perturbative
      regularization constructed}, it would be easier to prove the
      existence of the theory in the limit that the regulator is
      removed than in the ordinary, non-supersymmetric case.
      However, all known non-perturbative regularizations of gauge
      theories, including lattice methods, break supersymmetry.
      Hence, so far, we do not have at our disposal any
      non-perturbative definition of $N=4$ supersymmetric Yang-Mills
      theory\footnote{Recently it has been suggested that the
      correspondence just discussed may suffer from
      a fermion doubling problem\cite{doublingproblem}, and that
      this implies either a failure of the correspondence or 
      a breaking of supersymmetry in the
      $N=4$ supersymmetric Yang-Mills theory beyond perturbation theory.
      My understanding is that this issue is presently unresolved.}

      Since the theory itself has not yet been defined
      non-perturbatively, it follows that it cannot be used to provide a
      non-perturbative definition of another theory.  Thus, while it
      may be that in the future this situation is improved, for the
      time being it is not true that the Maldacena conjecture
      has been proved. Nor can it be said that the problem of giving a
      background
      independent or non-perturbative definition of string theory has
      been solved by assuming the conjecture is true and therefore
      defining string theory non-perturbatively in terms of
      the $N=4$ supersymmetric Yang-Mills theory.

  \subsection{Open issues of  string theory}

  It is clear from the summary just given that there is very good
  reason to take string theory seriously.  The theory appears to give
  a good perturbative description of quantum gravity through at least
  the two loop level and, even if this is true also of supergravity
  theories, this is still a very impressive fact.
  Many of the  results are very impressive, including the ones
  which show that there exist analogous systems in string theory with
  the same entropies and temperatures of extremal and non-extremal
  black holes.

  At the same time, there are a large number of open issues.

  \begin{itemize}

      \item{}There are a very large number of string theory
      backgrounds, labeled by both discrete topological classes
      and continuous parameters.

     \item{}So far, no string theory background is known which 
     is consistent with all features of the observed universe.
      They all have one or more of
the following features, which each disagree with observation:
no positive cosmological constant, unbroken supersymmetry, massless scalar
      fields.

      \item{}So far no string theory background is known which is 
      time dependent, as is our universe. Further no stable string theory 
      background is known which is
      consistent with recent observations that strongly suggest that
      there is a positive cosmological
      constant\cite{stringcosmoproblem}.

\item{}Further, we observe no massless scalar fields. Thus
in  any string background corresponding to
      nature there can be no such fields.  This means that the
      compactified geometry must be a consistent background only for
      discrete values of its parameters. At the same time, those
      parameters must have very small ratios in them, to explain the
      hierarchy problem.

      \item{}Even if a string theory background is found which is
also  consistent with everything that is observed, does this tell us anything,
      given that there is an infinite space of possible string
      backgrounds to search?  The theory would be predictive only if
      there were a unique string background consistent with what is
      observed. Is there any reason to believe this is the case,
      rather than there being a large or infinite number of such
      backgrounds\footnote{For estimates on the number of 
      non-supersymmetric string vacua, see
\cite{lenny-anthro,mikedouglas}. While no stable,
consistent, non-supersymmetric string vacua have yet been constructed, 
these authors argue, using recent developments, that the number of
such theories may be absolutely enormous.}?

      So, we should ask, even if there is a unique string theory
      background consistent with what is observed, how would nature
      pick it out? One might hope that there were a principle of
      stability or lowest energy that would pick out a unique
      string theory background. However this is unfortunately
      unlikely.
      We have good reason to believe\footnote{From
      results that support the conjecture of $S$-duality.} that many
of the
      supersymmetric vacua are stable. So it appears very unlikely that the
      observed background is the only stable one.

    Thus, even if string theory is true,
      there is so far no reason to believe that nature has a unique choice
      as to low energy phenomenology. Whatever the hopes, present
      evidence from string theory is more compatible with the idea
      that the observed background is picked out from many possible
      consistent ones by some dynamical process, occurring in the
early  universe, or even before the big bang\footnote{One such theory
      is cosmological natural selection\cite{CNS,LOTC}, which was invented
      to address
      this issue in string theory.  It remains falsifiable, but so
far  it has not been falsified and is
      consistent with all observations to date.}.

      This circumstance suggests that perhaps some attention be given
      to what might be called the search question in string theory.
      Given what we know, it is likely that if string theory is true,
      the real world is described by one out of a very large number of
      local minima of some potential or energy functional.
      This may be either a function on the space of string
      backgrounds or the expectation value of a potential or
      Hamiltonian in some fundamental Hilbert space of string theory.
      In either case, we have to find the global minimum of a function on
      a high dimensional space, which parameterizes possible string
      backgrounds. Further, we expect that the function has many
      local extrema, corresponding to the perturbatively consistent
      string theories. How do we find its global extremum?

      Further, if the global minimum is to agree with
      observed physics, there can be no
      massless moduli fields. This means that
      the true minimum must be an isolated point in the space of
      consistent string theories. However, all known
      consistent string theories have massless scalar fields, this
      means that all the local minima that correspond to them
      live in continuous submanifolds of solutions.

      How then are we to find the one true, isolated minima of
      a very complicated potential, which we know has lots of other
      minima, many of which have much more measure than the solution
      we seek?

      It is fair to ask whether examining them one by one, as they 
      are discovered by
      people putting together ever more complicated combinations of branes,
      orbifolds and complex manifolds is likely to hit on the true
      one. After all, the number of consistent backgrounds vastly
      outnumbers the number of people working in the field.
      Should we  be concerned that picking out the true minimum of a complex
      potential with a large number of local minimum is known from
      results in the theory of computation to be a
      very hard problem?

      A striking result of complexity theory is somewhat worrying in
      this regard. Called the ``no free lunch theorem'' this states that
      no specific search algorithm is likely to do better than random search in
      finding the global minimum of a randomly chosen complicated
      potential\cite{lunchcosts}. To do better than random search, a
      search procedure must be based on an algorithm which is crafted
      taking into account some properties of a given potential.

      This suggests that if we are ever to find the string vacua that
      describes our world we need to craft a search algorithm based on
     some non-trivial property of string theory, rather than just
     studying more and more complicated string vacua as the tools are
     developed to define them.

      \item{}It must also be emphasized that string theory does not
      give a genuine quantum theory of gravity, in the sense that
      each consistent string theory is defined with respect to a fixed,
      classical, non-dynamical background. So it is not background
      independent and it fails to address
      many of the questions that a quantum theory of gravity
      must answer\footnote{Some string theorists have argued that the
      Maldacena conjecture offers a non-perturbative definition of
      string theory that is background independent in the sense that
      all observables of quantum gravity are mapped into observables
      of the supersymmetric Yang-Mills theory. The only restriction
      is that the quantum spacetime must be asymptotically
      $AdS$.  This is perhaps an attractive proposal, but it itself depends
      on a positive resolution of some of the open issues discussed 
      above.}.

      Related to this is the fact that the black hole results in
      string theory do not concern actual black holes. They concern
      instead ensembles of states in free string theory in flat
      spacetime, with the
      gravitational coupling turned off. So far, no way has so
      far been found to extend the results to black holes that are
      not extremal or near extremal\footnote{Although there are
      suggestive results in a matrix form of string theory that this
      might become possible\cite{matrixbh}.  }.

      \end{itemize}

      To address all these issues,  a number of conjectures have been made.
  Some of these date back to the early days of string theory, 1984-5,
  and have been outstanding ever since. Several others have been
  added more recently.

  \subsection{Open conjectures of string theory}

  We have already mentioned four conjectures in string theory, so far
  unproven. These were

  \begin{itemize}
      
      \item{}{\bf Perturbative superstring theory is finite, unique
      and consistent to all orders in the genus expansion.}

      \item{}{\bf The black hole conjecture}

      \item{}{\bf The $S$ duality conjecture}

      \item{}{\bf The Maldacena conjecture}

\end{itemize}

Other conjectures which are believed by many string theorists include,

  \begin{itemize}

      \item{}{\bf Uniqueness of the non-perturbative ground state
      conjecture.}
      There is a unique  ground state in string theory, which is
      the solution of some dynamical
      problem, such as minimization of some potential.

      \item{}{\bf Empirical adequacy conjecture.} That unique string
      theory ground state leads uniquely to a prediction that the
      world has $3+1$ large dimensions, in which supersymmetry
      is broken, leading to
      a phenomenology in agreement with all
      observations\footnote{Alternatively, if the
      world is found experimentally to have
      more than $3+1$ uncompactified
      dimensions, as in the large extra dimension or
      Randall-Sundrum models\cite{RS}, this will be the
      unique prediction of the unique string theory ground state.}.

      \end{itemize}

      For there to be a dynamical mechanism to find a unique
      background, all the different backgrounds must be part of the
      same theory. So this requires:

      \begin{itemize}

      \item{}{\bf String theory unification conjecture.}
      The different background dependent theories are actually
      expansions around different classical solutions of a single,
      unified, background independent string theory, and this theory
      has a connected configuration space.

      \end{itemize}

      To support this conjecture a number of additional conjectures
      can be made. These generally depend on the $S$ duality
      conjecture. Indeed, if
      that conjecture is true, then
      it can be argued that if one takes
      together all the existing $S$ and $T$ dualities that
       all $5$ of the distinct string theories in flat ten
       dimensional spacetime are different descriptions of a single
       theory.

       Additional arguments suggest that this unified  theory
       also has backgrounds which are $11$ dimensional. This is
       because $10$ dimensional string backgrounds have in addition to
       the metric a scalar field. There is some evidence that this
       scalar field acts like a radius of an additional compactified
       dimension\cite{Mfirst,M-witten}.
       While there is apparently no consistent string theory
       in $11$ dimensions, there is a supersymmetric
       theory of gravity, called $11$ dimensional
       supergravity\cite{11sgrav}.

       There have then been discovered evidence for
       duality transformations that, at least approximately,
       relate certain features of $10$ dimensional string theories to
        $11$ dimensional supergravity\cite{Mfirst,M-witten}.
    There is also another interesting theory in $11$ dimensions,
    which is a description of a $2+1$ dimensional membrane,
    moving in $11$ dimensional spacetime\cite{dWHN}. This theory also is
    supersymmetric.   However,  it is not yet
    known if it has a consistent quantization.
    This leads to the

       \begin{itemize}

      \item{} {\bf $\cal M$ theory conjecture\cite{Mfirst,M-witten}.}
       There is a background independent formulation of string
       theory which unifies all the known string theories,  $11$
       dimensional supergravity and the $11$ dimensional supermembrane
       theory.

       \end{itemize}

       In some versions of the $\cal M$ theory conjecture the
       fundamental degrees of freedom are not strings in a ten
       dimensional spacetime. They are the three dimensional
       {\it membranes}, together with their duals, which are certain six
       manifolds, called {\it five-branes}, all existing in $11$
       dimensions. The idea is that $2$ dimensional string worldsheets
       are approximations to configurations in which one dimension of
       a membrane curls around one dimension of spacetime, and the
       radius of that dimension is taken very small. On larger scales,
       where spacetime seems $10$ dimensional, one sees only a two
       dimensional string.

       The beautiful thing about this conjecture is that it offers a
       possibility of an explanation of the $S$-duality conjecture
       in string theory.
       The different string theories which are conjectured to be
       related by $S$ duality arise, at least classically, from
       different ways of wrapping the added dimension of the membrane
       around the added dimension of space.
       Thus,  if there were an independent definition of $\cal M$ theory,
       or at least a consistent definition of the quantum membrane theory
       in $11$ dimensions, one might be able to prove the $S$ duality
       conjecture.  However, at present there is neither, so $S$
       duality, as well as $\cal M$ theory, remain interesting, but
       so far, unproven, conjectures.

       There are of course, various pieces of evidence that have been
       adduced for
       these conjectures.  Some of them are simply consequences of
       the symmetries, and can be explained by an understanding of
       how representations of the $11$ dimensional supersymmetry
       algebra decompose into representations of its $10$ dimensional
       sub-superalgebras. While beautiful mathematically, these hold
       whether or not there are
       consistent, quantum theories that realize the dynamics of the
       objects postulated.

       Some interesting developments that led to somewhat stronger
       results followed a conjecture that the dynamics
       of $\cal M$ theory can be formulated as a certain matrix
       model\cite{BFSS,IKKT}.
       This led to some non-trivial calculations of properties that
       $\cal M$ theory, were it to exist, would have. However, the
       matrix model was found to be strongly background dependent.
       It appears only to exist, or at least be tractable,
       in a limited set of backgrounds, mainly $11$ dimensional
       Minkowski spacetime and certain low dimensional toriodal
       compactifications of it.  Moreover, even in these cases, it
       was restricted to a certain limit, associated with light cone
       coordinates.

       Some work inspired by this has gone into attempts to extend
       this method to the construction of a genuinely background
       independent matrix model for $\cal M$
       theory\cite{stringsfrom}-\cite{latestmcs},
       but the results
       are not considered definitive.  In the absence of such a 
       formulation, there is no clear proposal for either the
       principles or mathematical formulation of $\cal M$ theory.
       It remains an interesting conjecture about the existence
       of a theory we do not so far know how to formulate or 
       construct.

I ended the section on loop quantum gravity by indicating how the 
approach is most likely to fail.  Some of the ways that string theory 
could fail, given present kowledge, include,

\begin{itemize}
    
    \item{}String theory could fail if there turn out to be no 
    consistent and stable\footnote{It might be argued that this could 
    be weakened to allow metastable vacua that are stable for times 
    long compared to the observed Hubble time.} 
    string vacua consistent with all the
    observed features of our universe including complete supersymmetry 
    breaking, the absence of massless scalar fields and a positive 
    cosmological constant.
    
    \item{}Conversely, string theory could fail if it turns out that 
    there are so many consistent and stable string vacua consistent 
    with all observations to date that they populate the space of 
    post-standard model physics densely enough that the theory makes 
    no predictions for future experiments. 
    
    \item{}String theory could also fail for theoretical reasons. For 
    example, it may turn out that it lacks 
    both a  perturbative defintion, if perturbative finiteness fails 
    past genus two, and a 
    complete non-perturbative definition (if, for example, all 
    attempts to construct non-perturbative regularizations of 
    supersymmetric Yang-Mills and string theories are subject to 
    fermion doubling problems that break supersymmetry.) 
    
\end{itemize}

It is also possible that string theory could pass these tests, 
but one or more of the open conjectures could fail, leading to a 
different physical picture than is widely believed.  For example, we 
may  note that the present evidence is consistent with the following
  pessimistic conjecture.

    \begin{itemize}

       \item{}{\bf Minimal string theory conjecture.}
        String theory only exists as a large number of background
        dependent theories. Perturbative superstring theory is not
	defined unambiguously or is not finite past genus two. 
	The various $S$ dualities that have been
        postulated do not in fact extend beyond the $BPS$ sectors and
        the different background dependent theories are not isomorphic.
    There is no connected configuration space and no unified theory that
    all perturbative string theories represent expansions around. 
    The conformal induction conjecture of Witten is true, but string 
    theory on $AdS^5 \times S^5$ and $N=4$ supersymmetric Yang Mills
    theory are inequivalent beyond that correspondence, so that all results
    to date involving gauge theory/supergravity or gauge theory/string theory
    correspondences are consequences of either conformal induction or
    the supersymmetry algebra applied to the $BPS$ states. Nor does
    string theory give a consistent description of quantum black hole
    spacetimes, apart from results concerning $BPS$ and near $BPS$ states.

  \end{itemize}

  Of course, most string theorists will be sure that this conjecture is 
  much 
  too pessimistic. I mention it only to emphasize the distance between 
  the picture often presented and assumed in many talks and 
  papers\footnote{For example, only one\cite{marshakov} out of 
  fifteen general
  review papers I consulted mentions that the question of whether
  superstring perturbation theory is finite and unambiguous to 
  all orders in the genus expansion 
  is unsolved.} on 
  string theory and the actual results to date.

  Even if some of this minimal conjecture turns out to be true, 
  the results that are on the table are among the most impressive
  and far reaching ever achieved in mathematical or theoretical 
  physics. So string theorists have a lot to be proud of. Even
  if part or all of the minimal conjecture turns out to be true
  there has
  been, and will remain a great deal to be learned from string
  theory, that may very well be relevant for 
  physics.

  Of course I do not know which of these conjectures will turn out to 
  be true. The fairest thing to say about string theory is that it is
  already a very impressive construction of mathematical physics, but 
  that the possibility of its relevance for a theory of nature depends 
  on substantial progress being made on the open conjectures listed 
  here.

\section{Other approaches}

Before summarizing our findings we should examine some of the other 
approaches which have been proposed to address the problem of quantum 
gravity.

\subsection{Dynamical triangulation models}

These are models in which a quantum spacetime is represented by
a simplicial complex. The edge lengths of the elements are fixed, and 
the degree of freedom is only the way in which a large number of
elements are connected together to make a simplicial complex. Each
element is assumed to model a region of spacetime on the order of
the bare Planck volume. 

Two classes of models have been extensively studied, both 
 numerically and analytically.  These are the Euclidean dynamical 
 triangulations models\cite{dynamical} and the causal
 dynamical triangulation models\cite{AL}.  
 
 \subsubsection*{Results from Euclidean dynamical triangulation models}
 
 In $2$ dimensions, the dynamical triangulation models are equivalent 
 to random surface models, and also equivalent to Louiville field 
 theory.  These models are completely solved, and all methods agree.
 
 In $4$ dimensions dynamical triangulation models were studied for
 several years\cite{dynamical}. The model has two phases and much work 
 was put into 
 determining whether the phase transition between them is first order 
 or second order. Were it second order it would make it possible to 
 show that the low energy limit of the model is general relativity. In 
 fact after much effort it was concluded that the phase transition is
 first order, so that general relativity is not a low 
 energy limit of the model.  This situation is believed by some 
 workers to persist when matter is added, although there are some 
 results to the contrary\cite{contrary-dynamical}.  Modulo the 
 resolution of this controversy, it is possible that this 
 approach to quantum gravity is ruled out.  
 
  \subsubsection*{Results from Lorentzian dynamical triangulation models}
  
  Following this failure, and in part inspired by causal spin foam
  models\cite{F-foam}, a new class of discrete models was 
  investigated, which are dynamical triangulations models of 
  spacetimes with Lorentzian signature.  This study has led to very 
  significant results\cite{AL}-\cite{c=1} which include,
  
a) A solution
to the infamous conformal mode problem, demonstrating in detail,
that the Lorentzian path integral is well defined\cite{conformalresolve}. 
The fluctuations of the conformal
model do not, as conjectured by Hawking and collaborators on the 
basis of semiclassical arguments, cause the 
path integral to be unbounded. Instead, the 
fluctuations are controlled by the path integral measure sufficiently 
so that the path integral remains well defined.

b) In $1+1$ dimensional the critical behavior was found and was
discovered to be very different from that of the $2$ dimensional
Euclidean theory. 
For example the Hausdorf dimension is $2$, rather 
than $4$ as in the case of Euclidean theories. This tells us that
naive expectations that the path integral for Lorentzian quantum
gravity could be defined by a naive analytic continuation from the
Euclidean theory is false. 

c) A certain problem, relevant for making a background independent 
form of string theory was, surprisingly, solved\cite{c=1}.  This arose from a 
matrix approach to string theory, which was however found to fail
if the dimension of spacetime was above one.  This is called the
$c=1$ problem. The results of Ambjorn, Loll and collaborators 
suggest that this
problem may be  resolved if the strings are modeled by their version of
Lorentzian dynamical triangulations. Their results show that the
theory exists in higher dimensions and that there is a phase
transition when one goes above one dimensional. Further, above
one dimension, the effective dimension of the string is three
in the low energy limit (the Hausforff dimension). This may be 
considered to be evidence 
that a theory of membranes may be relevant for a background
independent form of string theory. 
 
\subsection{Regge calculus models}

These are discrete models of quantum spacetime in which a spacetime 
history is represented as simplicial triangulation with varying edge 
lengths. Rather than varying the triangulation with fixed edge 
lengths, as in the dynamical triangulation models, the triangulation 
is considered fixed and the edge lengths are varied. This was one of 
the first models of quantum spacetime to be studied, and it continues 
to be studied today. 

In three spacetime dimensions the model was constructed many years 
ago by Regge and Ponzano\cite{RP}. Although its significance was not 
appreciated for some years, it was in fact the first example known of 
a topological field theory, and it remains a paradigmatic example of 
such a model. Its quantum deformation (where the deformation parameter 
is, as in loop quantum gravity, inversely related to the cosmological
constant) is rigorously defined and yields non-trivial invariants of 
three manifolds, knots and graphs.

In $4$ spacetime dimensions the model has two phases, but, as in the 
dynamical triangulation case, the transition between them appears to 
be first order, so that no continuum limit is found\cite{regge-results}.

\subsection{Causal set models}

This is an approach to quantum gravity based on a few simple 
observations about the role of causal structure in lorentzian geometry.
The causal relations among events in a lorentzian constitute a partial
ordering of the events. Given the causal relations among the events of
a spacetime, the metric of that
spacetime can be reconstructed modulo a local conformal factor and 
modulo spacetime diffeomorphisms. This implies that causal structure
plus a volume element together have exactly the right amount of
physical information needed to reconstruct the diffeomorphism 
equivalence class of a spacetime.  

This suggests the following two hypotheses:

\begin{itemize}
    
    \item{}{\bf Weak causal set hypothesis} A quantum spacetime history
    provides a list of events $\cal E$, with a partial order,  $\cal 
    P$, representing their 
    causal relations. When that quantum spacetime has a semiclassical 
    description in terms of a manifold $\cal M$ and a 
    diffeomorphism equivalence class of Lorentzian metrics $g_{ab}$,
    then 1) the events of $\cal P$ (or coarse grained sets of them) 
    can be imbedded in $\cal M$, 2) the causal structure of $\cal P$ can be 
    imbedded, modulo
    some method of coarse graining, in that of $g_{ab}$ and 3) the volume
    measured by $\sqrt{det(g)}$ counts the number of events in each
    region of $\cal M$ given by the embedding.
    
    \item{}{\bf Strong causal set hypothesis.} At the most 
    fundamental level, a quantum spacetime history consists of nothing
    but a discrete set of events $\cal E$ together with their causal
    relations $\cal P$. 
    
\end{itemize}    

The weak causal set hypotheses has been proposed in connection with
causal spin foam models\cite{Fotini-Wheeler}, as the causal evolution
of a spin foam (defined in \cite{F-foam}) gives rise to a discrete set of
events with causal relations. The weak causal set hypothesis may then 
become a tool to be used in the derivation of the low energy limit of 
a causal spin foam.

The strong causal set hypothesis was previously 
proposed by Sorkin and collaborators\cite{causalsets} and has been 
under development since. Recent results take up the proposal in
\cite{fmls1} that directed percolation may play a role in the low
energy limit of quantum gravity, in order to propose a dynamics for causal 
set models based on percolation\cite{poset-perc}.  

The main problem the strong causal set hypothesis has to solve is to 
give a dynamics for causal sets such that it is natural that $3+1$ 
dimensional spacetimes emerge at low energies. Large, randomly generated causal
sets are known not to resemble the causal structure of any low 
dimensional manifold. 
There is some evidence 
that the recently proposed directed percolation dynamics have a 
continuum limit that may correspond to a low dimensional geometry.

Another issue to be resolved is that the matter degrees 
of freedom must also be derived from the causal set.  The problem of 
course is that the fundamental structure is postulated to be so 
simple, that essentially every feature of our world besides the fact 
that there are causal relations must be deduced dynamically from the
low energy limit of the theory.

Having said this, the causal set program can claim one success, which 
is a correct prediction of the order of magnitude of the cosmological
constant\cite{causal-cc}. This is quite striking, 
given that no other approach has
so far anything convincing to say on this crucial problem. 

\subsection{Twistor theory \cite{twistor}}

This program of research also takes the causal structure of spacetime 
as more primary than the metric structure. It is based on the 
construction of a ``dual space" to a lorentzian manifold, called
twistor space, consisting 
of all the null lines (or planes in the complexified case.)  The causal
relations of the manifold are translated into topological relations
among submanifolds of twistor space. Twistor theory has been very 
successful in the context of classical general relativity and field 
theory, where it has led to important results. Characteristic of these 
results is that field equations on a spacetime are translated into 
conditions of complex analyticity on the dual twistor space. For 
example the self-dual Einstein equations have been solved in closed 
form in terms of the complex deformations of twistor space. The 
problem of translating the full Einstein equations into twistor space 
remains open, but it is still being pursued. There are also related 
results showing that the structure of 
general relativistic spacetimes can be expressed in terms of the null
rays.

With regard to quantum gravity, twistor theorists, led by Roger 
Penrose, hypothesize that the 
structure of twistor space should be translated into quantum theory.  
This has yet to be carried out fully, but there are intriguing results
involving the quantization of fields in twistor space. 

Twistor theory is closely tied to loop quantum gravity, in that the 
same simplification of the Einstein's equations in terms of the 
properties of self-dual connections and self-dual two forms plays an 
essential role in both programs. There are also suggestions that 
twistor theory is relevant for 
supergravity and string theory\cite{supertwistor}.

\subsection{Non-commutative geometry}

This is a program, proposed originally by Connes\cite{alain}, which
has had much recent influence in quantum gravity and string theory.
The basic idea of the original program is to characterize a Euclidean 
manifold in physical, diffeomophism invariant observables, in 
terms of the spectrum of the Dirac operator on the manifold. 
Connes then showed that there are structures which can be 
characterized by operator algebras that satisfy certain axioms 
satisfied by the Dirac operator on a manifold, that are, however, not 
constructed as manifolds from sets of points. This gives rise to a 
generalization of differential geometry, called non-commutative 
geometry. Connes proposed that the standard model of particle physics 
can be understood elegantly in terms of such a non-commutative geometry. 

Because of the use of operator algebras, it is natural to associate 
non-commutative geometry with quantum geometry. However, it should be 
cautioned that in some applications, non-commutative geometry is 
classical, in the sense that the physical $\hbar =0$, the deformation 
parameter which signals that the non-commutative manifold is not an
ordinary manifold, is then not identified with $\hbar$. 

Nevertheless, there are also proposals for identifying the  
deformation parameter with $\hbar$, so that non-commutative geometry 
becomes a genuine model of 
quantum geometry\footnote{But see \cite{eli-problem}, for a recent no 
go theorem on a certain approach to non-commutative geometry as quantum
geometry.}.

Apart from its intrinsic study, 
non commutative geometry has had an influence on string theory, 
as there are classes of string backgrounds that involve 
non-commutative ($\hbar =0$) geometries in their construction.  In such
applications non-commutative manifolds are sometimes characterized by 
saying that the coordinates have a non-commutative algebra. This is 
useful, but it should be mentioned that it is somewhat against the 
spirit of Connes' original approach, which was formulated in 
coordinate free, language in terms of diffeomorphism invariants.  

The quantum geometry discovered in loop quantum gravity is also 
(slightly) non commutative, in that operators that measure the areas 
and volumes of regions of a spatial manifold, strictly speaking fail 
to commute, although the lack of commutivity is only evident in their
action on a small set of states\cite{thomas-thesis}. 

\subsection{Condensed matter physics inspired models}

Recently several condensed matter physicists have proposed that quantum 
spacetime may be modeled in terms of an ordinary quantum statistical 
system such as a fermi liquid\cite{condensed,laughlin,zhang}. 
The idea is that even though such a 
system is defined with respect to a fixed background metric, and is 
generally even formulated as a non-relativistic system, there exist 
phases in which the spectrum of low energy excitations resembles that 
of massless particles in Minkowski spacetime. In some cases, 
excitations of different spin have the same propagation velocity, so 
that the low energy physics may be described to some approximation in 
terms of a relativistic field theory.  Thus, it has been proposed 
that perhaps the experimental success 
of special relativity is due to the universe 
being in such a low temperature phase, of a system that is 
fundamentally non-relativistic.  It is further postulated that 
general relativity may to some approximation be a manifestation of 
the dependence of the apparent ``speed of light" in these systems on 
parameters such as density and temperature. 

This program, needless to say, challenges the basic assumptions that 
underlie both special and general relativity. To succeed, it must 
show that the excitations of a non-relativistic condensed matter 
system really can be described in terms of relativistic fields. It must
explain the emergence at low energies 
of gauge symmetries and diffeomorphism invariance.  Furthermore, such 
a program is very vulnerable to falsification,  as it most likely 
predicts
modifications of the energy-momentum relations in the context of 
the preferred frame scenario A, discussed above. As discussed there, 
there are already quite strict limitations on such theories. 

At the same time, such studies may be useful as they may shed some 
light on how quantum critical phenomena may play a role in the 
emergence of classical spacetime and quantum fields in the low energy
limit of a spin foam model. 

\subsection{Induced gravity and effective field theory models}

Sakharov proposed some time ago that the fundamental theory might 
have only matter fields, on a classical background, 
so that the full  Einstein action might then appear as a quantum correction
to the effective action\cite{sakharov}.  
This idea has been explored from time to 
time since\cite{inducedgravity}, 
and has recently been used to study black hole
entropy\cite{frolov}. 
If one takes this as a proposal for a fundamental theory 
than one is saying that 
the gravitatational field is not itself subject to 
quantization. This means that the fundamental theory has a mixed form 
in which a classical metric background is coupled to quantum matter 
fields. This has been argued to be inconsistent on the grounds that it 
both leads to problems with interpretation and that it leads to 
instabilities, as the full quantum effective action contains 
curvature squared terms that contain unstable modes. 

Alternatively one might consider this proposal within the point of 
view of effective field theory, as an approximation to some full 
quantum theory of gravity with a cutoff of less than the Planck energy 
imposed. This is a sensible thing to do and it turns out that one can 
study certain quantum corrections to the gravitational force in this 
framework\cite{donoghue}.

\subsection{Asymptotic safety}

It was conjectured long ago by Wilson\cite{safe-wilson} and 
Parisi\cite{safe-parisi} that a perturbatively
non-renormalizable quantum field theory could be still well defined
if its renormalization group has a non-trivial ultraviolet fixed point, at some
non-vanishing values of the couplings. The idea was taken over
to quantum gravity by Weinberg, who called it asymptotic 
safety\cite{safe-weinberg}. Some evidence for an non-trivial
uv fixed point was found in a $2+\epsilon$ expansion\cite{safe-2}
and in a large $N$ expansion, \cite{safe-me}. However, the fixed
point found appears 
to involve curvature squared terms, and hence bring with it a danger
of instabilities and violations of perturbative unitarity, so 
the idea was for a long time dormant. Recently it has
been revived in \cite{safe-reuter}.

\section{How well do the theories answer the questions?}

 \begin{quotation}
      
     {\it  ``A proof is a proof. What kind of a proof? It's a proof. A proof is 
     a proof. And when you have a good proof, it's because it's proven.''

     \blankline
     
      Jean Chretien, Prime Minister of Canada.  }
      
      \end{quotation}

Let us now summarize how well the theories do answering the
questions.  The status of each of the 24 questions we asked,
in string theory and loop quantum gravity,
is summarized in Table 1. (The other programs answer each one
or a few of the questions, but so far do not address as many as
string theory and loop quantum gravity.)
\begin{table}[tbp]
    \centering
    \caption{Summary of results.
    A=solved.
    B=partial results, or solved in some cases, open in others.
    C=in progress using known methods.
    ?= requires the invention of new, so far unknown methods.
    -=makes no claims to solve.}
   \begin{tabular}{|c|c|c|}
       \hline
       Question & String theory & Loop Quantum Gravity  \\
        \hline
       Quantum Gravity   &   &   \\
       \hline
        1. {\it GR and QM true or need modification?}  & A  & A  \\
       \hline
       2. {\it Describes nature at all scales?}  & B  & A  \\
       \hline
       3. {\it Describes quantum spacetime geometry?} & B  & A  \\
       \hline
       4.{\it BH entropy and temperature explained?}  & B  & A  \\
       \hline
       5. {\it Allows $\Lambda >0$?}   & ?  & A  \\
       \hline
       6. {\it Resolves singularities of GR?} & B  & B   \\
       \hline
       7. {\it Background independent?} & ? & A  \\
       \hline
       8.{\it  New predictions testable now?} & ?  & B  \\
       \hline
       9. {\it GR as low energy limit?} & A & B   \\
       \hline
       10. {\it Lorentz invariance kept or broken?} & A  & B
         \\
       \hline
       11.  {\it Sensible graviton scattering?} & B & C   \\
       \hline
       Cosmology &  &   \\
       \hline
       1. {\it Explains initial conditions?} & ? & C    \\
       \hline
       2. {\it Explains inflation?} & C & C   \\
       \hline
       3. {\it Does time continue before big bang?} & ? & A  \\
       \hline
       4. {\it Explains the dark matter and energy?} & ? & ?  \\
       \hline
       5. {\it Yields transplankian predictions?} & C   & C    \\
       \hline
       Unification of forces &  &   \\
       \hline
       1. {\it Unifies all interactions?} & A & -  \\
       \hline
       2. {\it Explains $SU(3)\times SU(2)\times U(1)$ and fermion reps?} & ? & -  \\
       \hline
        3. {\it Explains hierarchies of scales?} & ? & -  \\
       \hline
       4. {\it Explains values of standard model parameters?} & ? & -  \\
       \hline
       5. {\it Unique consistent theory?} & ? & -  \\
       \hline
        6. {\it Unique predictions for doable experiments?} & ? &
        B  \\
       \hline
        Foundational questions &  &   \\
     \hline
     1. {\it Resolves problem of time in QC?} & ? & C   \\
       \hline
       2. {\it Resolves puzzles of quantum cosmology?} & ? & C    \\
       \hline
        3. {\it Resolves the black hole information puzzle?} & C & C    \\
       \hline
   \end{tabular}
    \label{tbl:}
\end{table}

Let us take the questions in turn, beginning with the questions
about quantum gravity.

\subsection{Quantum gravity questions}

Loop quantum gravity gives positive and specific answers to each of the
first ten
questions concerning quantum gravity. Specifically, so far, for the
case of nonzero $\Lambda$ every question is answered positively,
including the existence of a good low energy limit.  And, so far,
no modification of either the principles of general relativity or
quantum theory appears to be required for the existence of a good
quantum theory of gravity.

With regard to the second and third question, a complete physical
picture of quantum geometry is provided by the theory that differs in
striking and specific ways from the classical theory of spacetime
geometry.

With regard to the fourth question, there is a microscopic
description of black hole horizons, which reproduces and explains the
Bekenstein entropy in terms of conventional coarse grained
description
of microstates, in this case microstates of the horizon degrees of
freedom. Furthermore, calculations lead to a derivation of the
Hawking
spectra, with computable corrections.

There is no problem with a positive cosmological constant, in fact
this is the best case for the theory as here we have simultaneously a
microscopic and semiclassical description in terms of a single exact
solution to the quantum constraints.  Further, the temperature and
entropy of deSitter spacetime are understood.

Recent results show that cosmological singularities are removed.
There are no results yet concerning black hole singularities.

The theory is fully background independent.

The present indications are that there are Planck scale
modifications in the realization of global lorentz invariance,
leading to predictions for physical effects that may be observable in
the present and certainly will be testable in the near future.

With respect to question 11, the situation is not satisfactory, in
that no calculations of the scattering of gravitons past the
classical approximation have been yet carried out in loop quantum gravity.
It can be hoped that progress can be made soon, at least in 
the case of non-zero cosmological constant.

Thus, we may summarize by saying that so far loop quantum gravity
provides an answer to the first ten questions concerning quantum
gravity.    While the $11$'th question remains unresolved, there is now work
in progress which has a realistic chance of addressing it.

Next, we consider how well string theory answers the
quantum gravity questions.

String theory offers a possible answer to the first question of how
gravitation and quantum theory are unified, which does not require
that the principles of general relativity and quantum theory be
exactly compatible.  There is also evidence that it may
provide a solution to question 11,
in the form of superstring perturbation theory, so long as that theory
can be shown to be finite and unambiguous to all orders in the genus
expansion. 
These are both strong
successes of the theory.

In certain restricted cases, string theory does provide an answer to
question 2. These are BPS states, where the existence of $T$ and $S$
duality
transformation allows quantum geometry for scales shorter than the
string scale to be described in a dual theory in terms of scales
larger than the string scale.  In some cases, no deviation from the
classical picture in which spacetime is continuous and smooth, are
seen.  In other cases, it appears that the classical picture of
spacetime geometry becomes replaced by a non-commutative, but still
classical, (in the sense of $\hbar \rightarrow 0$), spacetime
geometry.

It is not known whether these results extend to all states and
solutions of a string theory, beyond the restricted set of $BPS$
states where calculations and duality transformations can be
explicitly carried out.

With regard to question 4, there are striking results in the case of
systems with the same quantum numbers as extremal
and near extremal black holes.  These results extend even to the
computation of grey body factors, in exact agreement with the
semiclassical results.  These suggest, but do not
show,
that string theory may in the future give a detailed microscopic
description of quantum black holes. It is not known presently whether
these results extend to all black holes. If they do not it may be
that
these results are accidental, in that they are forced by the
supersymmetry algebra that, in the case of $BPS$ and near $BPS$
states, strongly constrains the spectrum and degeneracies of the
Hamiltonian.

String theory so far does not appear to incorporate deSitter spacetime
as a consistent background, and hence has trouble with a positive
cosmological constant.

There are results that indicate that various kinds of singularities
of classical general relativity can be removed by string theory.
These
however do not so far include either cosmological or black hole
singularities.

String theory is not background independent, and as such offers
nothing new concerning question 9\footnote{The few attempts to construct
truly background
independent formulations of string theory have, in my view,
been promising, but have not generated so far strong interest.
Besides the proposal in \cite{mcs,latestmcs}, there are
a few other approaches, for example an approach based on
an $11$ dimensional Chern-Simons theory\cite{m-piotr}.
While interesting, this proposal faces the serious difficulty
that $11d$ Chern-Simons theory has many local degrees of freedom whose
dynamics and canonical structure are very tricky to
untangle\cite{highercs,5cs,11d}}.

Finally, string theory gives a theory of graviton scattering which is
known to be unambiguous and finite to second order in perturbation 
theory.  

\subsection{Cosmological questions}

Next, we come to the cosmological questions.

Recent results in loop
quantum
gravity  shows some promise of answering the first three of the
questions.  In particular, there are results that indicate that
cosmological singularities are eliminated and the evolution of the
quantum universe continues in time through the singularities of
classical $FRW$ universes\cite{LQC}. Calculations concerning transplankian
effects are in progress.  But loop quantum gravity has nothing to say about
the dark matter and energy.

String theory has so far little to say definitively about cosmological
questions. There are a number of ideas and models under development
to address those questions, using higher dimensions or the idea that
the universe lives on a brane in a higher dimensional manifold. Some
of these ideas are closely tied to string theory, others are not. It
is not yet clear whether this direction will lead to experimental
tests of string theory, but it is a possibility.  Very recently there
have begun attempts to compute transplankian effects using string
theory.

\subsection{Questions concerning unification}

Loop quantum gravity has, so far, nothing to say about the
question of unification. While there are some speculations in this
direction\footnote{Two proposals for unification within loop
quantum gravity are described in \cite{louisunify1} and
\cite{louisunify2} showing that this is a possibility that
merits further exploration.}, 
and some work on background independent approaches to
string theory, none of this has led to any definitive progress on the
questions asked.  However, as mentioned, it does appear that loop
quantum gravity makes predictions for experiments that test lorentz
invariance at high energies.

String theory was invented to be a unified theory of all the
interactions and its main strength remains the fact that it gives
a beautiful
and,
to many, compelling solution to the first question.
This is a great success, such theories do not grow on
trees.

At the same time, in a certain sense they do, as there turn out to be
an infinite number of background dependent string theories, all of
which provide consistent perturbative unifications of gauge theories
with gravity, coupled to a variety of matter fields including 
fermions,
at least through genus two in perturbation theory.

As such, string theory so far provides no answer to the other
questions concerning unification. There is so far no known stable string
background that predicts all the observed features of particle
physics phenomenology, or resolves any of the open questions
concerning the standard model of particle physics. String theory
makes so far three clear predictions: 1)  supersymmetry
should be found at some scale between a Tev and the Planck scale.
2) at some scale some evidence for additional dimensions must be
found, 3) exact lorentz invariance should be preserved.

All three of these are the object of current experimental programs.
One should note that supersymmetry and evidence of additional
dimensions or degrees of freedom need not be discovered in near
future experiments, as the only absolute prediction is they
must be discovered somewhere below the Planck scale. Nor
would the discovery of Tev scale
supersymmetry
prove the correctness of string theory, as there are ordinary
supersymmetric field theories which are extensions of the standard
model.  Nor would the discovery of supersymmetry rule out loop
quantum gravity, as all the main results of loop quantum gravity
can be extended to supergravity, at least through $N=2$.

It is then evident that present and near future tests for the
presence of Planck scale corrections to the energy momentum relations
offer both the best way to falsify string theory and the best way to
distinguish experimentally between string theory and loop quantum
gravity.

Further, given the existence of an infinite number of string theory
backgrounds, in the absence of a complete non-perturbative
formulation
of string theory, it seems that there is not on the horizon any way
to strongly confirm unique predictions of string theory, as opposed to
predictions of supersymmetric grand unified theories. For example, it
is compatible with all known results that, if there is any string
theory background consistent with all known experiments, there are a
large and perhaps infinite number of such models, that would give
different predictions for phenomenology at Tev and higher scales. In
this case string theory would still fail to be predictive, beyond the
general prediction that Lorentz invariance should be realized linearly
at all scales.

Finally, it should be emphasized that so far neither string theory
nor loop quantum gravity have much new to offer to resolve the
questions of the large hierarchy of scales. They are both compatible
with various mechanisms proposed to resolve the gauge hierarchy problems
in grand unified theories, as there are
versions of each that incorporate the basic features of (possibly
supersymmetric) grand unified theories. 
Both theories also appear to be compatible with known field
theoretic mechanisms for the spontaneous breaking of supersymmetry.
Neither theory has so far
anything to offer to explain why the cosmological constant is so
small, although at least loop quantum gravity appears to have no
problem with the apparently observed fact that the sign is positive.

\subsection{Foundational questions}

This leaves the last two questions, concerning the problem of time
and the problem of quantum theory in a closed universe.  Since loop
quantum gravity provides a precisely defined example of a quantum
theory of gravity and cosmology these problems may now be
investigated with a precision that was not previously possible.
The result, to summarize a lot of work, is that most of the so far
proposed
solutions to the problem of time in quantum cosmology
can be formulated in detail and tested
in the context of loop quantum gravity.   There is presently a lively
debate concerning this issue, so I will not try here to predict the
outcome, except to say that there appears to be no reason to believe
the problem is no more difficult in principle than that in the full
classical theory,
with cosmological boundary conditions.  That is, if one sticks
strictly to discussing physical, gauge invariant observables, and is
careful to ask only physical questions, than the different notions of
time which have proved useful in the classical theory can be
constructed and
represented in the quantum theory.

With regard to the question of the formulation of a measurement theory
for quantum cosmology, when the observer is part of the universe,
much
the same situation obtains. The different possible solutions which
have been proposed can be expressed exactly in the Hilbert space of
solutions to the Hamiltonian constraint of loop quantum gravity.

While this is an area of lively debate, it may be noted that a new
kind of solution to the problem of providing a measurement theory to
quantum cosmology has been formulated by people working in loop
quantum gravity\cite{louis-holo,pluralism,carlo-relate,Fotini-QCH,BI}.
This is called {\it relational quantum cosmology}
and the basic physical ideas are due to Crane, Rovelli and Markopoulou.
A mathematical structure for a generalization of quantum theory
that appears compatible with their ideas has been proposed
by Butterfield and Isham\cite{BI}.

The basic ideas of relational quantum cosmology are described in an 
appendix.

Last but not least we come to the black hole information puzzle.  While it is
fair to say that to date niether the results of 
string theory nor of loop quantum gravity, {\it per 
se}, resolve this problem, both research programs have given rise 
to claims about how it may be resolved. {\it If the strong Maldacena 
conjecture is true}, then, at least in the asymptotically $AdS$ case,
one can argue that there can be no loss of information because whatever
goes on in the bulk, it can be represented by unitary evolution in the
dual gauge theory, which is an ordinary quantum field theory in flat 
spacetime.  While this claim has been made by a number of authors,  
it must be emphasized 
that, as we have seen above, the evidence so far supports far more 
strongly the weaker conjecture of conformal induction than it does the
strong Maldacena conjecture.   If only conformal induction holds than 
there still could be loss of information in bulk black hole 
evaporation without contradicting the correspondence between a 
restricted set of boundary
observables in the bulk theory and the boundary theory. 

In particular, one possible resolution of the black hole information 
puzzle has always been that black hole singularities bounce to create
new universes, which expand to the causal future of all points on the
horizon of the black hole\cite{LOTC}. In this case the causal 
structure of the spacetime implies there is a permanent loss of 
information from the point of view of an external observer, as some 
information is only accessible to an observer who falls into the black 
hole and goes through the bounce.  Were there a real proof of the 
Maldacena conjecture it might imply that this scenario does not 
occur, at least in asymptotically $AdS$ black holes. However, the 
present results certainly do not rule it out.

Such a resolution is entirely unproblematic from the point of view of 
relational quantum theory, as that requires only a local conservation 
of information. As described in \cite{algebraic-fotini}, this 
is realized naturally 
by weakening the requirement of global unitarity to a local 
condition, which is that evolution is described by completely positive maps. 
From the point of view of relational quantum theory, the general
situation is that the causal structure restricts the generic physical 
observer inside a universe to have access to less information than 
would be required to reconstruct a pure state for the universe. The
problem of loss of information in black hole evaporation is then just
one example of a more general situation. This situation is resolved 
by reformulated quantum theory entirely in terms of density matrices
representing the partial information accessible to real, physical, 
local observers embedded in a spacetime.

\section{Conclusions}

The first thing that must be 
said is that if we compare what we know now about quantum gravity to what we
knew twenty years ago, it is clear that there has been enormous
progress.  This is due to an enormous effort by a large number of
people, who choose to dedicate their time and, in many cases, their
careers to pursue this very risky venture, when they might have found
success more quickly elsewhere.  It is impossible to look at the list
of results in these and other approaches to quantum gravity over the
last twenty years and not feel in awe of the enormous talent,
intelligence and hard work that people have contributed.

Second, both string theory and loop quantum gravity are very much alive as
research programs with a significant chance of uncovering new laws of
nature. Each has achieved much more than prudent
experts would have bet was possible twenty years ago. Certainly more
progress has been made on quantum gravity than I expected to
see in my lifetime.  So it is clear that both loop quantum gravity and
string theory should continue to be pursued vigorously.  Each
deserves significant support  from the
physics and academic communities.

The same may be said for several of the other approaches to quantum
gravity. For example, the recent progress in lorentzian dynamical
triangulation models gives us the first firm results about a number
of key issues in quantum gravity such as the role of Euclideanization
and the conformal mode problem. Nor would it be surprising if other
appraoches such as non-commutative geometry, twistor theory or
causal sets play a key role in the final theory of quantum gravity. 

At the same time, my own conclusion after the excercise of writing 
this
review is that the different theories are in very 
different situations. To explain this impression I would like to 
propose a list of what
would need to be done in each case to finish the theoretical program
and bring each theory to the point where it could be compared with
real, doable, experiments. Any such list requires a certain amount of
speculation and guess work, and I am sure that different people would
produce, if asked, different lists. But it is interesting nonetheless to get 
an idea of what remains to do in each case.

\subsection{What remains to be done in loop quantum gravity?}

\begin{enumerate}

    \item{}For general $\Lambda$ develop the method of coherent 
    states to discover the conditions for a loop quantum gravity
    theory to have a consistent low energy limit, develop
    perturbation theory around it and test whether it is 
    consistent to all orders.

    \item{}For the case $\Lambda \neq 0$ develop perturbation theory
    for excitations around the Kodama state and test whether it is
    consistent and sensible order by order.

    For general $\Lambda$, continue the development of renormalization group
    methods in the context of spin foams, to discover which loop quantum
    gravity theories have good low energy limits.   

    \item{}Develop a method in loop quantum cosmology to predict
    transplankian effects in cosmology and, when they are accessible
    to test, compare them to precision measurements of $CMB$ spectra.
    
     \item{}Refine the existing calculations that predict modified
    energy-momentum relations, to determine whether or not the theory
    makes unique predictions
    for the parameters $\alpha, \beta \ldots$ in the modified
    energy-momentum relations (\ref{modified}), 
    for the different particle species.   Resolve the question of
    whether Lorentz invariance is realized exacctly, 
    broken or realized non-linearly
    in the low energy limit of loop quantum gravity.  
    
    \item{}Develop a dynamical formulation of spacetimes with 
    horizons in order to understand dynamically 
    the connection between the discreteness of area
    and the quasi normal mode spectrum discovered by Dreyer. 
    
    \item{}Work out the deatils of a 
    version of the holographic principle in the 
    context of relational quantum cosmology, making use of the fact 
    that the Bekenstein bound is realized naturally in constrained
    topological field theories. 

\end{enumerate}

\subsection{What remains to be done in string theory?}

\begin{enumerate}
    
    \item{}Resolve the problem of the existence, uniqueness and
    consistency of superstring perturbation theory past genus two. 

    \item{}Demonstrate the existence of at least one string perturbation theory
    consistent with all the features of our universe, including
    completely broken supersymmetry, a positive cosmological constant
    and the absence of massless or light fundamental scalar fields.

    \item{}Discover whether this theory, if it exists, is unique and,
    if so, whether any predictions can be made for near term experiments.

    \item{}Related to the foregoing, understand how and why
    supersymmetry, if present at all, is spontaneously broken.

    \item{}Discover whether the $S$ duality conjecture is true or false.

    \item{}Discover whether the Maldacena conjecture is true or false.

    \item{}Find a background independent formulation of string or
    $\cal M$ theory. Find the classical solutions to this theory and
    show that the different perturbative string theories do arise as
    expansions around them.

    \item{}Formulate a principle that picks out a unique perturbative
    string theory which may be our universe.   Explain why this principle
    picks a universe with all the features of ours, including having $3+1$
    ordinary, large dimensions, broken supersymmetry, no massless
    scalars, and whose low energy phenomenology is given by the
    standard model.

    \item{}Follow this with calculations of the parameters of the
    standard model of particles physics and, perhaps, cosmology.

    \item{}Make unique predictions for phenomena that are beyond the
    predictions of the standard model, but accessible to present or
    near term experiments.
    
    \item{}Fully develop a version of the holographic principle in 
    string theory, either by proving the Maldacena conjecture or
    else by finding an alternative formulation. Show that it applies 
    also to cosmological spacetimes with horizons. 

    \item{}Find methods to study general black hole spacetimes in
    string theory.

    \item{}Develop an approach to cosmological singularities in
    string theory. Then develop a method to extract predictions for
    transplankian effects and compare them to future $CMB$
    observations.

    \item{}Finally, give a simple set of postulates for string theory
    from which all the results relevant for the description of nature
    may be derived.

\end{enumerate}

Different people might propose different items for such a list. In mine
there is a clear difference which emerges, which brings out the
differences in the two resarch programs. 

This is due perhaps to the fact that string theory is a far more ambitious
program than loop quantum gravity. String theory is perhaps
best understood to be 
a research program in search of new postulates for fundamental 
physics, 
whereas loop quantum gravity is based on the combination of the
relatively well understood principles of quantum theory and general
relativity. As a result, loop quantum gravity is perhaps less ambitous,
but because of this it appears to be significantly closer to completion. 
After a long period of development during which
results have accumulated, the claims that could be made for loop
quantum gravity have steadily strengthened.  At this point the theory
is well enough understood that it is possible to formulate a
program to bring it to completion and experimental test over the next
several years, using only known ideas and methods.

What remains
to be done on the theory side 
requires mostly the application of standard methods such as
perturbation theory and renormalization group techniques to well 
defined theories. Regarding 
cosmology there is a research program under development, using
standard methods, that is likely to result in predictions for
transplankian phenomena, that may be testable.  For
$\Lambda \neq 0$ there are good indications that the 
difficult problem of showing that general
relativity is the low energy limit may be solved, and
perturbation theory is presently under development.
Even for $\Lambda =0$, where the Kodama
state does not guarantee the existence of a good low energy limit, there are
techniques under development which, while computationally
challenging, should allow
increasing control over the low energy limit.

Meanwhile,  a set of experiments that may allow the theory
to be tested have been identified and calculations are underway to
sharpen the predictions for them. These require only the application
of standard methods of theoretical physics. 

No new principles are required because 
loop quantum gravity I takes as its postulates only the
principles of general relativity and quantum theory, and these have
the great advantage of being well confirmed experimentally.  What the
results tell us is that, when due attention is paid to issues of
how to incorporate the gauge invariances of general relativity in a
quantum field theory, there appears no obstacle to the joint
application of these well established sets of principles.  What
remains may be no more than the straightforward working out
of the consequences for experiment.

Loop quantum gravity I may fail. For example, the predictions of quantum
general relativity may turn out to disagree with experiment.  The good
news is that this may occur within the next ten years. Even if it 
does, loop quantum gravity II offers a conceptual and mathematical
framework for a large class of quantum theories of space and time. It 
is then not a specific theory, it is more analogous to lattice gauge 
theory in being a general technique to investigate theories with 
certain kinds of symmetries, in this case diffeomorphism invariance. 
It can even be said that  
loop quantum gravity II offers a framework for the ambition motivating 
string theory to be realized that is not hampered by background dependence 
and is based on a complete and exact unification of quantum theory with
the basic principle that space and time are fully dynamical and
background independent. 
But even loop quantum gravity II may turn out to make generic 
predictions that disagree with experiments to be carried out in the
next decade. 

By comparison, the situation of string theory is much less
clear, at least for the near future.  One problem is that
several of the steps on the list 
remain unsolved, after many attempts over many 
years. It is possible then that these will
require the discovery of  new, presently
unknown, principles  and,
quite possibly, also substantial mathematical and technical innovations.
As what is needed goes significantly beyond what is known,
it seems not easy to predict when, or over what path,
string theory may be able to take the steps necessary to become a
completed physical theory. 

Nevertheless, progress is continuing. One question on which there are
new approaches, if not yet a solution, is the second,  that of finding a 
string perturbation theory that is
not ruled out as realistic by some observed feature of the 
world\footnote{Very recently proposals have been made for string 
vacua that have positive cosmological constants and break 
supersymmetry\cite{positive-string}.These theories are all unstable 
and, it is argued, must decay.  There are indications that some of
these theories may have decay times long compared to the present Hubble
scale, but it is not yet clear that all decay channels have been 
understood.}.  It
is possible that this may be achieved by continuing to construct
models with existing methods. At the same time, it is
consistent with present results to conjecture, as does
Banks\cite{N-Banks},
that there are no
consistent string perturbation theories that admit a positive
cosmological constant, as presently observed, or  that have
completely broken supersymmetry with no massless scalar fields.
At the very least, it appears possible that
string theories with these characteristics, if they exist at all, may
require the introduction of new methods.

As an indication of this situation, it is interesting that 
recently a few string theorists have been
resorting to mention of the anthropic principle in order to
resolve the problem of connecting the theory with observation.
Recently Susskind\cite{lenny-anthro} 
and others have proposed that there
will be a huge number of consistent non-supersymmetric 
string vacua (although still not one stable non-supersymmetric
string vacua has so far been written down) and that there will be no
selection mechanism to pick out a preferred one except the
anthropic principle\footnote{For a detailed 
discussion of exactly what kinds of ``anthropic reasoning" do and
don't lead to theories that satisfy the basic test of a scientific 
theory that it be falsifiable, see \cite{LOTC}. As described there in
detail, most
versions of the anthropic principle fail to be falsifiable. An
example of a genuinely falsifiable theory (which incidentally has
so far survived attempts to falsify it) is cosmological natural
selection, described also in \cite{LOTC}.}.  This appears to 
represent quite a change
of perspective from claims that string theory in the end would
lead to a unique recovery of standard model physics and testable predictions 
for observations beyond the standard model.  This is another 
indication that string theory is a search for new principles. 

These difference are certainly  due to the fact that the 
two programs have very different 
ambitions. String theory began as a search for a conjectured unique 
theory that would unify all of nature. In spite of the fact that 
at the background dependent level string theory is far from unique,
to a large extent this is still the prime motivation of the program. 
To realize this hope, 
string theory relies on several mathematical conjectures which remain
unproven, in spite of intense effort, 
and several physical hypotheses, which may turn out to be
right or wrong.  While the idea of duality, that gauge and other 
degrees of freedom may be described in terms of stringlike 
excitations,  is attractive, 
the cost of realizing it as a fundamental, rather than an effective 
theory, appears high. Either there are or are not extra dimensions, and
supersymmetry is either part of the laws of nature or not. In the
end only experiment can tell, but there appears to be no near term experimental
program which could falsify these hyptheses. 
What is so frustrating about string 
theory is that it could easily be wrong, in whole or in part, but
there appear to be few realistic ways to find out. The only 
possibility I know of for near term falsification 
are the tests of lorentz invariance at high energies. 

On the other hand loop quantum gravity makes much less radical
assumptions, and instead investigates the question of {\it how}
to fully reconcile the basic physical ideas and principles that
underlie relativity and quantum theory.  While still incomplete, loop
quantum gravity has clearly succeeded in partly solving this
problem, resulting in several novel physical predictions, 
and it is the only research program that has done so. String
theory has so far largely ignored the problems loop quantum gravity
has taken on, and to a large extent solved.

If string theory is right then sooner or later it will have to attack
the problem of how to have a fully consistent background
independent quantum theory of space
and time.  It will then have to begin to address the issues that loop
quantum gravity has already gone a long way towards  solving.

From the point of view of loop quantum gravity II, supersymmetry and
higher dimensions can be easily incorporated and there is also
good reason to believe that strings may emerge as an effective
description at a scale below the Planck scale\cite{stringsfrom}.  
Thus, from the point
of view of loop quantum gravity, if experiment shows that the world is
supersymmetric or higher dimensional, there need be no obstacle
to describing it in background independent terms.
Thus it remains a possibility that in the end string theory
and loop quantum gravity will come together because the
methods and results of loop quantum gravity will turn out to be 
indispensable for
the solution of the problem of making a background independent form of
string theory.

The most important conclusion of this survey is that there
is now a realistic chance that experiment may over the next ten years
be able to distinguish between the predictions of different quantum
theories of gravity, including string theory and loop quantum gravity.
Given this, the first priority of theory must be to anticipate the
experiments, by bringing the theories to the point where they make
clean predictions that may allow them to be falsified.

\section*{ACKNOWELDGEMENTS}

This review started as an essay commissioned in honour of John 
Wheeler, and although it was not finished in time for that I would
like to thank the editors of the Wheeler volume and the Templeton 
foundation for the invitation to write it. I also have to 
thank many people for discussions and correspondence 
over the last several years
that were helpful to its preparation. 
These include Giovanni Amelino Camelia, 
Tom Banks, John Brodie, Shyamolie Chaudhuri,
Michael Douglas, Willy Fischler,
Michael Green, Brian Greene,
David Gross, Murat Gunyadin, Gary Horowitz, Chris Hull,
Chris Isham, 
Clifford Johnson,
Renata Kallosh,  Jurek Kowalski-Glikman, 
Frederic Leblond, Juan Maldacena, David Mateos, 
Djorge Minic,
Rob Myers, Herman Nicolai, Amanda Peet, Joe Polchinski,Lisa Randall, Konstatin
Saviddy, Steve Shenker, Kelle Stelle, Andy Strominger, Raman Sundrum, 
Lenny Susskind, 
Thomas Thiemann, Arkady Tseytlin
and Edward Witten.   I am especially grateful to Eric D'Howker, Clifford
Johnson, Gordy Kane, Juan Maldacena, Rob Myers, John Schwarz,
and Arkady Tseytlin for taking the time to answer questions that came
up while writing this review, to Seth Major for very helpful comments
on the text and to Fotini Markopoulou for very helpful critical advice.
I also must thank my collaborators over the
years who have taught me most of what I know about quantum gravity,
Stephon Alexander, Matthias Arnsdorf, Abhay Ashtekar,
Roumen Borissov, Louis Crane,  John Dell,
Ted Jacobson, Laurent Friedel, Yi Ling, Seth Major,
Joao Magueijo, Fotini Markopoulou, Carlo Rovelli and Chopin Soo. I
am grateful as well to
the NSF and the Phillips Foundation for their very generous support which
made my own work possible.  Finally I would like to thank my 
friends and colleagues at Perimeter Institute for the open,  critical, 
but friendly atmosphere which was so helpful in preparing this review 
and especially to Frederic Leblond for an inviation to present
this work at the PI string seminar and for many helpful and critical
remarks there. 

\section*{APPENDIX: Relational Quantum Cosmology}

For the interested reader, I summarize here the
basic ideas of relational quantum cosmology.

\begin{itemize}

\item{}Crane:  Hilbert spaces are associated with boundaries that
split the universe into parts.  By the relationship of GR to TQFT
these will be described in terms of
{\it finite dimensional state spaces}\cite{louis-holo}.

\item{}Rovelli:   The Hilbert space describes the information one
part of the universe has about another part\cite{carlo-relate}.

\item{}Markopoulou:  Different Hilbert spaces are associated with
local observers inside the universe and describe information coming
from their past light cones. Quantum cosmology without the
wave function of the universe\cite{Fotini-QCH}.

\item{}Butterfield and Isham: The right mathematics for relational
quantum theory is topos theory\cite{BI}.

\end{itemize}

Reduced to a slogan, relational quantum cosmology maintains that,
``Many quantum states to describe one universe, not one state
describing
many universes."

In fact relational quantum cosmology is closely connected to the
holographic principle. Indeed, its original formulation, by Crane,
preceded 't Hoofts papers on the holographic principle, and should
probably be considered the first statement of the principle.

One version of the holographic principle connected with relational
quantum cosmology has been proposed, which may be called the
{\it weak holographic principle.} It may be summarized as saying
that\cite{weakholo}

\begin{itemize}

\item{}A surface in space is a channel through which quantum
information flows.  All measurements are made on such surfaces.
Each surface has associated to
it a Hilbert space that contains the possible outcomes of measurements
made on that surface.

\item{}The area of the surface is another name for its capacity as a
channel of quantum information. The log of the dimension of the Hilbert
space of each surface is hence taken to be a definition of its area. In
this way geometry is reduced ultimately to information theory.

\item{}This is the basis of a measurement theory for spatially closed
causal spin foam

\end{itemize}

In fact, as conjectured by Crane in the paper that stimulated these
developments\cite{louis-holo}, in loop quantum gravity there are Hilbert spaces
associated with boundaries and they are constructed from Chern-Simons
theory, which is a topological field theory.  Moreover they are,
as Crane conjectured, finite dimensional and they do automatically
implement Bekenstein's bound. Thus, while there remains work to do to
fully formulate  a relational quantum cosmological theory, it may be said that loop
quantum gravity does have some features suggested by relational
quantum theory.

Relational quantum theory has also been explored in a few papers by
string theorists, particularly Banks and Fischler\cite{BF-relate}.  They
propose that when the cosmological constant, $\Lambda >0$,
supersymmetry is necessarily broken and the
quantum theory of gravity is described in terms of finite dimensional
Hilbert spaces.  They then propose an approach that appears to have
some elements in common with that of 
relational quantum cosmology, particularly in the form given by
Markopoulou\cite{Fotini-QCH}.

One consequence of these ideas is that there is, at least when
$\Lambda >0$, no need for Hilbert spaces
that appear in the theory to be infinite dimensional. Instead, one
can argue that  $N$, the dimension of
any local Hilbert space arising in the theory is bounded by
$G \Lambda \over 3$.  This bound was conjectured by
Banks\cite{N-Banks} on the basis
of the fact that this is the entropy of deSitter spacetime.  

Thus, the conclusion is that new ideas have arisen in loop quantum
gravity which have some hope of resolving the problem of time and the
problem of quantum cosmology.  Further, as a well formulated
background independent quantum theory, loop quantum gravity allows
older ideas about these problems to be precisely formulated and tested.
Meanwhile, while string theory apparently does not  offer so far
anything new to resolve these problems, it is striking that a few
string theorists have put forward proposals that appear to be inspired
by the ideas coming from loop quantum gravity.

  \end{document}